\documentclass[12pt]{article}
\usepackage{amsfonts}
\usepackage{bm}
\usepackage{amsmath}
\usepackage{epsfig}
\usepackage{slashed}

\usepackage{amssymb,lscape}
\usepackage[matrix,arrow,curve]{xy}

\newcommand{\bmat}{\left(\begin{array}}
\newcommand{\emat}{\end{array}\right)}

\def\gtrsim{\mathrel{\raise.3ex\hbox{$>$\kern-.75em\lower1ex\hbox{$\sim$}}}}

\def\p{\partial}

\def\G{\Gamma}
\def\d{\delta}

\def\L{\Lambda}

\def\Om{\Omega}

\def\-{\hphantom{-}}

\def\s2{\frac{1}{\sqrt2}}

\def\D{{\Delta}}

\def\mg{m_{3/2}}
\def\mg2{m^2_{3/2}}

\def\Dsl{\,\raise.15ex\hbox{/}\mkern-13.5mu D} %this one can be subscripted

\def\be{\begin{equation}}
\def\ee{\end{equation}}
\def\bea{\begin{eqnarray}}
\def\eea{\end{eqnarray}}

\newcommand{\nn}{\nonumber}

\newcommand{\cC}{{\cal C}}
\newcommand{\cD}{{\cal D}}
\newcommand{\cE}{{\cal E}}
\newcommand{\cF}{{\cal F}}
\newcommand{\cG}{{\cal G}}
\newcommand{\cH}{{\cal H}}
\newcommand{\cJ}{{\cal J}}
\newcommand{\cN}{{\cal N}}
\newcommand{\cR}{{\cal R}}
\newcommand{\cZ}{{\cal Z}}

\newcommand{\dop}{{\slashed\nabla}}

\reversemarginpar

\begin{document}

\pagestyle{plain}

%----------------------------------------------------------------------%
%  numbering equations with section number
%----------------------------------------------------------------------%
\makeatletter
\@addtoreset{equation}{section}
\makeatother
\renewcommand{\theequation}{\thesection.\arabic{equation}}
%----------------------------------------------------------------------%
%  title page
%----------------------------------------------------------------------%
\pagestyle{empty}
 \begin{center}
 \ \ \
 \\[6mm]

 {\LARGE \bf Exploring Double Field Theory }\\[6mm]
 \vskip 1cm

 {\bf David Geissb\"uhler$^{1}$,\, Diego Marqu\'es$^{2}$,\, Carmen N\'u\~nez$^{2,3}$ \,and\, Victor Penas$^{3}$}\\

\vskip 25pt

 {\em $^1$ Albert Einstein Center for Fundamental Physics,\\
Institute for Theoretical Physics, University of Bern }\\

 \vskip 0.3cm

 {\em $^2$ Instituto de Astronom\'ia y F\'isica del Espacio (CONICET-UBA)}

 \vskip 0.3cm

 {\em $^3$ Departamento de F\'isica, Universidad de Buenos Aires (CONICET-UBA)}

\vskip 0.8cm

 \end{center}

 \vskip 1cm

 \begin{center}

 {\bf ABSTRACT}\\[3ex]

 \begin{minipage}{13cm}
 \small

We consider a flux formulation of Double Field Theory, in which geometric and non-geometric fluxes are dynamical and field-dependent. Gauge consistency imposes a set of quadratic constraints on the dynamical fluxes, which can be solved by truly double configurations. The constraints are related to generalized Bianchi Identities for (non-)geometric fluxes in the double space, sourced by (exotic) branes. Following previous constructions, we then obtain generalized connections, torsion and curvatures compatible with the consistency conditions. The strong constraint-violating terms needed to make contact with gauged supergravities containing duality orbits of non-geometric fluxes,
systematically arise in this formulation.

\end{minipage}
\end{center}

\newpage
%----------------------------------------------------------------------%
%  Resetting of counters
%----------------------------------------------------------------------%
\setcounter{page}{1}
\pagestyle{plain}
\renewcommand{\thefootnote}{\arabic{footnote}}
\setcounter{footnote}{0}
%----------------------------------------------------------------------%
%  Paper begins
%----------------------------------------------------------------------%

\tableofcontents

\section{Introduction}

String dualities reveal intriguing relations among perturbatively different theories.
While T-duality establishes the physical equivalence of theories defined on dual backgrounds with very different geometries,
S-duality relates the strong and weak coupling limits
 of dual theories, and finally U-duality has been conjectured to be a symmetry of the full string theory.

Much progress has been achieved in the construction of duality covariant models
aiming at an effective description of the low-energy states of the string, their interactions and properties.
The stringy nature of the dualities alters the standard notions of geometry, and in  some of the
approaches  duality invariance is achieved
through an enlargement of the coordinate space.
The idea of implementing T-duality as a manifest symmetry was first considered by M. Duff \cite{duff} and A. Tseytlin \cite{tseytlin} and further developed by W. Siegel
\cite{Siegel:1993th}. More recently, it received renewed attention after the works by C. Hull, B. Zwiebach and O. Hohm \cite{Hull:2009mi},
where the theory defined on the doubled space
was named Double Field Theory (DFT)  (see also \cite{doublegeom}).
The equivalence between the formulations in \cite{Siegel:1993th} and \cite{Hull:2009mi}
 was established in \cite{DFTgeomHohmKwak}.
Closely related is the framework of Generalized Geometry
\cite{Hitchin},
\cite{Gualtieri:2003dx}.
More general U-duality covariant frameworks have been constructed in \cite{Uduality},\cite{Aldazabal:2013mya} and \cite{E11programme}, and
 the relation between some of these theories and DFT was explained in \cite{Thompson}. A review of these achievements can be found in \cite{amn}

DFT is usually supplemented {\it ad hoc} with a differential constraint on fields and gauge parameters, named {\it strong constraint} or {\it section condition}. It effectively un-doubles the double coordinate dependence, and implies that locally DFT is a reformulation of supergravity. Given the coordinates of the double space $X^M$, $M = 1,\dots,2D$, and the corresponding derivatives
$\partial_M=\partial /\partial X^M$, the constraint states that
 \be \eta^{MN}\partial_M \partial_N \dots =  0 \ ,  \ \ \ \ \ \eta^{MN} = \left(\begin{matrix}0 & \delta_i{}^j \\ \delta^i{}_j & 0\end{matrix}\right) \, ,\label{strongconstraint}
 \ee
 where $\eta^{MN}$ is the $O(D,D)$ invariant metric, $i,j = 1, \dots , D$ and the dots stand for arbitrary (products of) fields and gauge parameters.  Generalized diffeomorphisms in the double-space then reduce to standard diffeomorphisms and two-form gauge transformations.

 The first step towards a relaxation of the strong constraint was implemented in the Ramond-Ramond sector \cite{Massive}. For the Neveu-Schwarz sector, it was shown in \cite{Aldazabal:2011nj} that closure of the algebra of generalized diffeomorphisms and gauge invariance of the action of DFT give rise to a set of constraints that are not in one to one correspondence with the strong constraint. Although they  imply that DFT is a restricted theory,  solutions that violate the strong constraint and are thus truly doubled are allowed.

Scherk-Schwarz (SS) compactifications \cite{ss} provide a scenario where fields and gauge parameters are restricted: given a background defined by a duality twist, the fields and gauge parameters must {\it accommodate} to it, and can no longer be generic. The perturbations around the background then correspond to the dynamical degrees of freedom of the effective action, which is a gauged supergravity.
When the restricted fields are inserted into the consistency constraints of DFT, the duality twist generates gaugings (including the so-called non-geometric gaugings \cite{Shelton:2005cf}) that arrange in the form of the quadratic constraints of gauged supergravities  \cite{Aldazabal:2011nj}. Then, under a SS reduction, the constraints of DFT are in one to one correspondence with the constraints of gauged supergravity. U-duality invariant scenarios exhibit the same behavior \cite{SS Uduality}, \cite{Aldazabal:2013mya}. The quadratic constraints were completely solved in some particular gauged supergravities in \cite{Dibitetto:2012rk}, where it was shown that the  duality orbits of non-geometric fluxes  are only generated through truly doubled duality twists. From a phenomenological point of view, these duality orbits (which necessarily violate the strong constraint) are the most interesting ones since they favour moduli stabilization and dS vacua, evading the many no-go theorems for geometric fluxes \cite{Danielsson:2012by}. Then, from a four-dimensional perspective, the effect of the strong constraint is to eliminate the orbits that give rise to vacua with desirable phenomenological features.

The purpose of this paper is to explore to what extent one can deal with the gauge consistency constraints in DFT
 without  imposing  the strong constraint, and survey extensions of DFT with strong constraint-violating terms. To achieve this goal, we closely follow the formulation in \cite{Siegel:1993th}, \cite{DFTgeomHohmKwak}:
\begin{itemize}
\item The fields of the theory, namely the generalized dilaton $d(X)$ and bein ${\cal E}_A{}^M(X)$, which turns flat indices $A, B, \dots$ into
curved ones $M,N,\dots$,
 are arranged in ``dynamical'' fluxes defined as:
    \begin{eqnarray}
{\cal F}_{ABC} &=& 3 \Omega_{[ABC]}\ , \\
{\cal F}_A &=& \Omega^B{}_{BA} + 2 {\cal D}_A d\, ,
\end{eqnarray}
where
\be
\Omega_{ABC} = {\cal D}_{A} {\cal E}_B{}{}^N {\cal E}_{CN} \, ,\ee
and
we have introduced a planar derivative ${\cal D}_A = {\cal E}_A{}^M \partial_M$.
The fluxes ${\cal F}_{ABC}$ and ${\cal F}_A$ are thus field-dependent and non-constant.
The different components of ${\cal F}_{ABC}$ correspond to the standard geometric ($H_{abc}$ and $\tau_{ab}{}^c$) and non-geometric ($Q_a{}^{bc}$ and $R^{abc}$) fluxes, and give rise to the corresponding gaugings upon compactification. This is similar to the constructions of \cite{Andriot:2011uh}, where  ten-dimensional actions with their associated differential geometries were built in terms of field dependent quantities related to the non-geometric fluxes.

\item Some consistency constraints take the form of generalized quadratic constraints, and involve the following Bianchi identities (BI) for the dynamical fluxes
\be\label{Constraints}
 \begin{split}
   \cD_{[A}{\cal F}_{BCD]}-\frac{3}{4}{\cal F}_{[AB}{}^E {\cal F}_{CD]E} &= {\cal Z}_{ABCD}\ ,\\
    \cD^C {\cal F}_{CAB} +2\cD_{[A}{\cal F}_{B]}-{\cal F}^C {\cal F}_{CAB}&= {\cal Z}_{AB}\ ,
\end{split}
\ee
where
\be\label{const2}
\begin{split}
    {\cal Z}_{ABCD} \equiv& -\frac{3}{4}\Omega_{E[AB}\Omega^E{}_{CD]}\, ,\\
{\cal Z}_{AB} \equiv& \left(\partial^M\partial_M {\cal E}_{[A}{}^N \right){\cal E}_{B]N}
                    - 2\Omega^C{}_{AB} \cD_C d\, .
\end{split}
\ee
Upon SS compactifications, the constraints lead to the quadratic constraints for the constant electric bosonic gaugings of half-maximal gauged supergravity.
Both these expressions vanish under the strong constraint (\ref{strongconstraint}), but more generally
the full set of constraints admits truly double configurations.  Let us emphasize that the strong constraint can be imposed on all the results of this paper, which would then reduce to known results in the literature.

Besides (\ref{Constraints}) there are additional BI associated to the quadratic constraints of the maximal theory, which arise upon completing the NS-NS action  with the Ramond-Ramond (RR) sector
\be
\cD^A{\cal F}_A-\frac{1}{2}{\cal F}^A{\cal F}_A +\frac{1}{12}{\cal F}^{ABC}{\cal F}_{ABC}=-2\cD^Ad\,\cD_A d+2\partial^M\partial_M d+\frac{1}{4}\Omega^{ABC}\Omega_{ABC} \equiv {\cal Z}\nn
\ee
\be
	\dop \cG = \cZ_{RR}\ ,
\ee
where $\cal G$ contains the information on RR forms, and $\dop$ is a generalized Dirac operator. All ${\cal Z}...$ vanish under the strong constraint.

\item The action takes the form of the scalar potential of the bosonic electric sector of
half-maximal gauged supergravity \cite{Schon:2006kz} when the fluxes are identified with the constant electric gaugings and the flat metric is identified with the modular scalar matrix:
\bea
\!\!\!\!\!\!\!\!\!\!\!\!\!\!\!\!\!\!\!\!\!\!\!\! S \!\!\!\! &=&\!\!\!\! \int dX e^{-2d} \bigg(\!\!- \frac 1 4 {\cal F}_{AD}{}^C {\cal F}_{BC}{}^D S^{AB} - \frac 1 {12} {\cal F}_{AC}{}^E {\cal F}_{BD}{}^F S^{AB} S^{CD} S_{EF} + \cF_A\cF_B S^{AB}  \nn \\
&&\ \ \ \ \ \ \ \ \ \ \ \ \ \  -\frac{1}{6}\cF^{ABC}\cF_{ABC}- \cF^A\cF_A \bigg)\, ,\label{actionintro}
\eea
where $S_{AB}$ is the generalized metric in planar indices, and it is written purely in terms of the dynamical fluxes.
Up to boundary terms, the first line in this action equals that of DFT \cite{Hull:2009mi} plus an additional term that violates the strong constraint. The second line, on the other hand, identically vanishes
under  the strong constraint. So, when the strong constraint is imposed, this action reduces to the action of the generalized metric formulation of
DFT \cite{Hull:2009mi}.
These results refer to the NS-NS sector, but we also include Ramond-Ramond fields and heterotic vectors in the analysis.
\end{itemize}

The action (\ref{actionintro}) includes many strong constraint-violating terms, some of which were added {\it by hand} in \cite{Aldazabal:2011nj}, and of course were absent in the original formulations of DFT. These terms are covariant under the global and local symmetries, up to the quadratic constraints, and are needed to make contact with
half-maximal gauged supergravities containing duality orbits of non-geometric fluxes in four-dimensions upon a SS compactification.
Here, we construct this action systematically as in \cite{Siegel:1993th} closely following the guidelines of \cite{Geom} (and also \cite{DFTgeomHohmKwak},\cite{Park},\cite{WaldramR}): we first introduce connections to covariantize the derivatives under the gauge symmetries of the theory and then impose a set of conditions on them, such as vanishing generalized torsion and compatibility with  the dynamical degrees of freedom  and the $O(D,D)$  metric. Although only some projections of the connection are determined, a notion of generalized Riemann tensor can be introduced which, upon traces and projections, leads to a fully determined generalized Ricci tensor (whose flatness determines the equations of motion) and  a generalized Ricci scalar (that defines the action (\ref{actionintro})). This procedure is followed here without assuming the strong constraint (this was also done in the U-duality case in \cite{Aldazabal:2013mya}, and also in a different geometric construction of DFT \cite{BermanGEOM}). We find that the strong constraint-violating terms appearing in the generalized Ricci scalar are those introduced in \cite{Aldazabal:2011nj} plus others that are needed to guarantee gauge invariance (the latter play no role when a SS compactification is performed) up to the consistency constraints.

Let us emphasize that in this paper we don't assume a SS form of fields and gauge parameters: we simply list  the consistency constraints of the theory that appear  through the computations, and show that in particular they admit truly doubled solutions of the SS form. Other compactification scenarios might provide new solutions to the constraints.

Interestingly, the expressions (\ref{Constraints}) appear all along the many computations in the paper. They arise when analyzing closure of the gauge transformations, covariance of the generalized fluxes (which in turn implies gauge invariance of the action), invariance of the action under double Lorentz transformations, covariance of the generalized Riemann and Ricci tensors, and they also show up in the BI for the generalized Riemann tensor.

 It is also interesting to note that when the strong constraint is imposed on the fields, (\ref{Constraints})  become the BI  of \cite{Shelton:2005cf},\cite{FreedWitten} for constant fluxes, and those of \cite{Blumenhagen:2012pc} for non-constant fluxes. They span T-duality orbits of BI, containing $\partial_{[i} H_{jkl]} = 0$  as a particular representative. These identities are known to be sourced by localized branes (see for example \cite{Villadoro:2007tb}), like the NS5-brane. More generally, we have here duality orbits of BI for non-geometric fluxes that can be related to more exotic T-fold-like objects with non-trivial monodromies, such as the $5^2_2$ brane \cite{deBoer:2010ud}, or other Q and R-branes \cite{Hassler:2013wsa}, etc. We also have duality orbits of generalized BI for branes in other dimensions and D-branes \cite{Doublebranes}, all related to the consistency constraints of DFT.

We stress that the formalism implemented here to analyze possible relaxations of the strong constraint was introduced in the pioneer work by W. Siegel \cite{Siegel:1993th} many years ago, and was recently extensively discussed in \cite{DFTgeomHohmKwak} by O. Hohm and S. Kwak. This includes the fluxes, action, BI, and other issues considered in this paper.

The paper is organized as follows. In Section \ref{ReviewDFT} we introduce the dynamical fluxes and, in terms of them, the action, equations of motion and gauge consistency constraints. In Section \ref{Geometry} the novel notions of stringy differential geometry are adapted to hold beyond the strong constraint. The inclusion of Ramond-Ramond fields and heterotic vectors is discussed in Section \ref{TypeIIhet}. In Section \ref{BianchiId}
we analyze the generalized BI, we present a first order formulation of DFT and discuss duality orbits of generalized BI for different types of branes. Finally we conclude and summarize in Section \ref{conclusions}.

%%%%%%%%%%%%%%%%%%%%%%%%%%%%%%%%%%%%%%%%%%%%%%%%%%%%%%%%%%%%%%%%%%%%%%%%%%%%%%%%%%%%%%%%%%%%%%%
\section{Double Field Theory and generalized fluxes}\label{ReviewDFT}

Double Field Theory is a manifestly T-duality invariant field theory
in which the fields depend on a double set of coordinates
 dual to momentum and winding.  Its simplest version contains only NS-NS fields, namely the metric $g_{ij}$, the antisymmetric Kalb-Ramond two-form $B_{ij}$ and the dilaton $\phi$. Extensions that include heterotic vector fields \cite{heterotic}, Yang-Mills symmetries \cite{Yang-mills}, R-R forms \cite{TypeII} and fermions in a supersymmetric fashion \cite{fermions} were also considered. The connection with $O(D,D)$ covariant world-sheet theories
was established in \cite{worldsheet}.

In its simplest version, the theory has a global symmetry group
$G = O(D,D)$,
 with metric
\be
 	\eta_{MN}
	= \left(\begin{matrix} 0 & \delta^i{}_j \\ \delta_i{}^j & 0 \end{matrix}\right)\ ,
	\ \ \ \ M,N = 1,\dots,2D \ , \ \ \ \ i,j = 1, \dots,D\, .
\ee
Curved indices $M,N,\dots$ are raised an lowered with this metric.
Every object appearing in a   duality invariant  theory must belong to some representation of the duality group $G$. In particular, the space-time coordinates $x^i$ have to be supplemented with $G$-dual coordinates $\tilde x_i$ to form generalized coordinates $X^M = (\tilde x_i, x^i)$, lying in the fundamental representation of $G$. It is in this sense that the theory is {\it doubled}. It also enjoys a gauge invariance generated by a pair of parameters $(\tilde \xi_i,\xi^i)$, that can be packed in the $G$-vector $\xi^M$. Gauge invariance and closure of the gauge algebra lead to a set of differential constraints that restrict the theory. In particular, these constraints are satisfied when a stronger condition named {\it strong constraint} is enforced:
\be
	\partial_M \partial^M \dots = 0 \, ,\label{strong}
\ee
where the dots denote (products of) fields and gauge parameters. The effect of (\ref{strong}) is to locally restrict the coordinate dependence of the fields and gauge parameters so that they live on a null $D$-dimensional subspace of the double space. In other words, when the strong constraint is imposed, the theory is not truly doubled but only lives on a $D$-dimensional slice of the doubled space. However, by explicitly breaking the gauge symmetry, for instance when compactifying, it is possible to relax the strong constraint. All through  this paper we will  keep terms that would vanish by this constraint.

In DFT, the  dilaton $\phi$ is contained in the $G$-scalar
\be
	d = \phi - \frac{1}{2} \log \sqrt{ g}\ ,
\ee
which is manifestly T-duality invariant. The $D$-dimensional metric $g_{ij}$ and two-form $B_{ij}$ are contained in a
symmetric generalized metric ${\cal H}_{MN}$, living in the coset $G/H$
where $H = O(1, D-1) \times O(1, D-1)$ is the maximal (pseudo-)compact
subgroup of $G$  corresponding to a local symmetry of the theory.
Therefore ${\cal H}_{MN}$ satisfies the constraint
\be
	\cH_{MP}\eta^{PQ}\cH_{QN} = \eta_{MN}\ .
\ee
A possible parameterization is the following
\be\label{stdparam}
	{\cal H}_{MN}
	= \left(\begin{matrix} g^{ij}  &  -g^{ik} B_{kj}\\
	 B_{ik} g^{kj}&  g_{ij} - B_{ik} g^{kl} B_{lj} \end{matrix}\right)\, .
\ee

Given these objects, an invariant action
under the gauge and  global transformations can be found, namely \cite{Hull:2009mi}
\be
S=\int d X e^{-2d}{\cal R}({\cal H}, d)\ ,\label{action1}
\ee
with
 \be\label{ActionDFTgenmetstrconst}
 \begin{split}
  {\cal R} \ \equiv  &~~~4\,{\cal H}^{MN}\partial_{M}\partial_{N}d
  -\partial_{M}\partial_{N}{\cal H}^{MN} + 4 \partial_M {\cal H}^{MN}  \,\partial_Nd  -4\,{\cal H}^{MN}\partial_{M}d\,\partial_{N}d
 \\[1.0ex]
    ~&-\frac{1}{2}{\cal H}^{MN}\partial_{M}{\cal H}^{KL}\,
  \partial_{K}{\cal H}_{NL} +\frac{1}{8}\,{\cal H}^{MN}\partial_{M}{\cal H}^{KL}\,
  \partial_{N}{\cal H}_{KL} + \Delta_{SC} {\cal R}\, ,
 \end{split}
 \ee
where $ \Delta_{SC} {\cal R}$ stands for terms that vanish under (\ref{strong})
and were not included in \cite{Hull:2009mi}.
This action reduces to the standard supergravity action for the NS-NS sector when ${\cal H}_{MN}$ is parameterized as in (\ref{stdparam}) and
the strong constraint (\ref{strong}) is enforced in a frame in which $\tilde \partial^i = 0$.

In the frame formulation of DFT, one takes the $H$-invariant metric as
\be
S_{AB} = \left( \begin{matrix} s^{ab} & 0 \\ 0 & s_{ab}\end{matrix}\right) \ , \ \ \ \ a,b = 1, \dots, D \ , \ \ \ \ s_{ab} = {\rm diag}(-+\dots +)\, .
\ee
When compared with standard supergravity, one of the $O(1, D-1)$ factors reproduces the local Lorentz symmetry. The generalized metric can then be written in terms of a generalized bein ${\cal E}^A{}_M$ as
\be
	{\cal H}_{MN} = {\cal E}^A{}_M S_{AB} {\cal E}^B{}_N\, .
\ee
A possible parameterization, leading to (\ref{stdparam}), is
\be\label{beinparam}
 	{\cal E}^A{}_M
	= \left(\begin{matrix}e_a{}^i & e_{a}{}^k B_{ki} \\ 0 & e^a{}_i\end{matrix}\right)\, ,
\ee
where $e^a{}_i$ is a $D$-dimensional bein of the metric $g_{ij} = e^a{}_i s_{ab} e^{b}{}_j$.

The indices in $H$ are always raised and lowered with the flat counterpart of the $G$-metric
\be\label{oddc}
	\eta_{AB} = \cE_A{}^M\cE_B{}^N\eta_{MN} =
	\left(\begin{matrix} 0 & \delta^a{}_b \\ \delta_a{}^b & 0 \end{matrix}\right)\, .
\ee
The last equality is verified by the parameterization (\ref{beinparam}), but for a generic doubled bein this gauge choice is a constraint forcing $\cE_A{}^M$ to be an element of $G$ itself. The additional degrees of freedom contained in the bein compared to those in $\cH_{MN}$ are then un-physical due to the new local symmetry $H$. Throughout this paper, we will generally not make use of any particular parameterization but rather consider the bein as a constrained field satisfying (\ref{oddc}).

Under global $G$ transformations, the generalized coordinates and fields transform as
\be
	X^M \to X'^M = g^M{}_N X^N \ , \ \ \ \
	{\cal E}_A{}^M (X) \to {\cal E}_A{}^N (X') g_N{}^M  \ , \ \ \ \
	d (X)\to d(X')\, ,
\ee
where $g \in G$ satisfies $\eta_{MN} = g_M{}^P \eta_{PQ} g_N{}^Q$. As mentioned above, when introducing beins, the theory enjoys a new Lorentz-like local symmetry $H=O(1, D-1) \times O(1, D-1)$ acting on $\cE^A{}_M$ from the left.
We note however that the constraint (\ref{oddc}), and all differential identities that follow from it, are invariant under {\it local} $G$
transformations (denoted $G_L \sim G$)
acting on the bein from the left
\be
	{\cal E}_A{}^M (X) \to h_A{}^B (X)\, {\cal E}_B{}^M (X)\ ,
\ee
where $h\in G_L$ satisfies $\eta_{AB} = h_A{}^C \eta_{CD} h_B{}^D$. The action and dynamical equations are however only invariant under the subgroup $H\subset G_L$, i.e. under transformations satisfying
in addition $S_{AB} = h_A{}^C S_{CD} h_B{}^D$.

%%%%%%%%%%%%%%%%%%%%%%%%%%%%%%%%%%%%%%%%%%%%%%%%%%%%%%%%%%%%%%%%%%%%%%%%%%%%%%%%%%%%%%%%%%%%%%%
\subsection{Flux formulation}

We would now like to rewrite DFT in terms of $G$-singlets only, along the lines of \cite{Siegel:1993th}. For this purpose we define the flat derivative $\cD_A = {\cal E}_A{}^M \partial_M$ and the {\it Weitzenb\"ock connection}
\begin{equation}\label{DynOmega}
	\Omega_{ABC}
	=  \cD_A{\cal
E}_{B}^{\ M}
 {\cal E}_{CM} = -\Omega_{ACB}  \, ,
\end{equation}
where the antisymmetry follows from (\ref{oddc}). Comparing compactifications of DFT with $\cN=D=4$ gauged supergravity, it was remarked in \cite{Aldazabal:2011nj} that the  objects:
\begin{eqnarray}
{\cal F}_{ABC} &=& 3 \Omega_{[ABC]}\, , \label{DynFabc}\\
{\cal F}_A &=& \Omega^B{}_{BA} + 2 \cD_A d\, , \label{DynFa}
\end{eqnarray}
play an important role. In particular, it was realized that
when they are constant and the indices refer to the internal group O(6, 6),
they can be identified with the electric gauging parameters
$f_{ABC}$ and $\xi_A$, or {\it fluxes} entering the embedding tensor. Moreover, the different components of these dynamical fluxes correspond to covariant derivatives of scalars, curvature of the gauge fields, and other covariant combinations that appear in the effective action.

The dynamics of the NS-NS sector of DFT
is described by an action that can be written in a compact form
(up to total derivatives) in terms of a scalar function of the generalized bein and dilaton as
\be\label{actionaction}
	S =\int dX\ e^{-2  d}\ {\cal R}({ {\cal E}}, d)\, ,
\ee
where
\bea
	\cR &=&  S^{AB} \left( 2\cD_A\cF_B - \cF_A\cF_B\right)
	+ {\cal F}_{ABC} {\cal F}_{DEF}
	\left[\frac{1}{4} S^{AD} \eta^{BE} \eta^{CF}-\frac{1}{12} S^{AD} S^{BE} S^{CF}\right] \nn\\
&& -2D^A\cF_A + \cF^A\cF_A -\frac{1}{6}\cF^{ABC}\cF_{ABC}\, .\label{R}
\eea
Here, the bein appears only through $\cD_A$, $\cF_{ABC}$ and $\cF_A$. When the parameterization (\ref{beinparam}) is chosen, and the strong constraint is imposed in the global frame in which the dual coordinate dependence vanishes, this action reduces to the usual NS-NS action of supergravity. Other parameterizations and global frames are better to describe the dynamics of non-geometry \cite{Andriot:2011uh}.

The second line in (\ref{R})
identically vanishes under the strong constraint. Up to boundary terms, the first line can be taken to the form of the standard action of DFT (\ref{ActionDFTgenmetstrconst}),
modulo a single strong constraint-violating term that was introduced in \cite{Aldazabal:2011nj}. It was also mentioned in \cite{Aldazabal:2011nj} that a term proportional to $\cF^{ABC}\cF_{ABC}$
should be added to the action (\ref{action1}) to recover the scalar potential of half-maximal gauged supergravity
in four dimensions. The second line in (\ref{R}) corresponds to the $H$ (and $G_L$, since it does not depend on the planar  generalized metric) invariant extension  of this term, up to the consistency constraints. When non-vanishing, its effect  for compactifications is to add a piece to the dilaton potential, which is indispensable to reproduce duality orbits of non-geometric fluxes.

Comparing (\ref{R}) with (\ref{ActionDFTgenmetstrconst}) we see that  the missing strong constraint-like terms read
\be
\Delta_{SC}{\cal R} =  \frac 12(S_{AB} - \eta_{AB}) \partial_M {\cal E}^A{}_P \partial^M {\cal E}^B{}_Q \eta^{PQ} + 4  \partial_M d \partial^M d - 4 \partial_M \partial^M d \ .\label{DeltaSCR}
\ee

The first line in
(\ref{R}) is also invariant under a $\mathbb{Z}_2$ symmetry reproducing the $B \to -B$ symmetry of supergravity.
This symmetry acts at the same time on the left and on the right of the bein by an $O(2D)$ transformation
\be
	\mathbb{Z} = \begin{pmatrix} \mathbb{I} & \\ & -\mathbb{I} \end{pmatrix}, \qquad
	\cE \to \mathbb{Z}\cE \mathbb{Z}\ .
\ee
Since $\mathbb{Z}\eta \mathbb{Z }= -\eta$, only terms involving an even number of contractions with $\eta$ are
invariant, and so is the first line in (\ref{R}).
The second line in (\ref{R}) instead breaks the $\mathbb{Z}_2$ symmetry.
It was shown in \cite{Aldazabal:2011nj}, based on the results of \cite{Aldazabal:2011yz}, that its presence forbids an embedding of the effective action
of DFT into $\cN=8$ supergravity in four dimensions. In order to truncate ${\cal N } = 8 \to 4$ in four-dimensions, a $\mathbb{Z}_2$ symmetry is imposed, and only the invariant terms are kept. It is therefore to be expected that such a symmetry is related to the one mentioned here.
Actually, let us mention that the quadratic constraints of gauged supergravities are automatically  solved by the
strong constraint (\ref{strong}). The second line in (\ref{R}) can be recast as
\be
{\cal Z}=\cD^A{\cal F}_A-\frac{1}{2}{\cal F}^A{\cal F}_A +\frac{1}{12}{\cal F}^{ABC}{\cal F}_{ABC}=-2\cD^Ad\,\cD_A d+2\partial^M\partial_M d+\frac{1}{4}\Omega^{ABC}\Omega_{ABC} \label{gf_ktens0}
\ee
and written in this way, it is easy to see that it vanishes under the strong constraint. In terms of $\cal Z$, (\ref{DeltaSCR}) can then be written as
\be
\Delta_{SC}{\cal R} =  \frac 1 2 S_{AB} \partial_M {\cal E}^A{}_P \partial^M {\cal E}^B{}_Q \eta^{PQ} - 2 {\cal Z} \ .
\ee
Given its relation to the quadratic constraints of maximal supergravity \cite{Aldazabal:2011nj}, ${\cal Z} = 0$ must not be imposed as a constraint here, unless we intend to embed DFT in some U-duality invariant theory. In this paper we will keep this term, which in fact allows for the possibility of obtaining duality orbits of non-geometric fluxes upon compactifications \cite{Dibitetto:2012rk}. Interestingly, when analyzing the RR sector of the theory, $\cal Z$ will appear as part of the consistency constraints.

%%%%%%%%%%%%%%%%%%%%%%%%%%%%%%%%%%%%%%%%%%%%%%%%%%%%%%%%%%%%%%%%%%%%%%%%%%%%%%%%%%%%%%%%%%%%%%%
\subsection{Gauge symmetries and constraints}
Under an infinitesimal $G_L$-transformation parameterized by $\L_A{}^B$, with $\L_{AB} = - \L_{BA}$, the bein transforms
as
\be
\d\cE_A{}^M = \L_A{}^B \cE_B{}^M\ .
\ee
Referring to definitions (\ref{DynOmega})$-$(\ref{DynFa}),
we then obtain the variations
\bea
\d_\Lambda\Om_{ABC} &=& \cD_A \L_{BC}
    + \L_A{}^D\Om_{DBC} + \L_{B}{}^D\Om_{ADC} + \L_{C}{}^D\Om_{ABD}\ ,\\
\d_\Lambda\cF_{ABC} &=& 3\left(\cD_{[A} \L_{BC]} +  \L_{[A}{}^D\cF_{BC]D}\right)\ ,\\
\d_\Lambda\cF_{A} &=& \cD^{B} \L_{BA} + \L_{A}{}^B\cF_{B}\ .
\eea

For $H$-transformations, the parameters also satisfy $\L_{\check{A}B} = \L_{A\check{B}}$, where we introduced the notation
\be
\L_{\check{A}B} = S_A{}^C\L_{CB} \ .\label{Check}
\ee
Then, up to boundary terms we find
\be
\delta_\Lambda S=  \int dX e^{-2d} \Lambda_A{}^C (\eta^{AB} - S^{AB}) {\cal Z}_{BC}\, , \label{VarHaction}
\ee
where
\be
{\cal Z}_{AB} = \cD^C {\cal F}_{CAB} +2\cD_{[A}{\cal F}_{B]}-{\cal F}^C {\cal F}_{CAB}=\left(\partial^M\partial_M {\cal E}_{[A}{}^N \right){\cal E}_{B]N}
                    - 2\Omega^C{}_{AB} \cD_C d\ .\label{gf_ktens2}
\ee
Notice that this vanishes under the strong constraint (\ref{strong}), but more generally $H$-invariance only requires the following minimal constraint
\be
(\delta_{[A}{}^C - S_{[A}{}^C) {\cal Z}_{B]C}=0\, . \label{gaugeinv2}
\ee
Here the $S$ contribution comes from the first line in (\ref{R}) and the $\eta$ term from the second line. Notice that invariance of the full action requires this projection of ${\cal Z}_{AB}$ to vanish, but if ${\cal Z}_{AB}$ is requested to vanish entirely as a constraint, then the action splits in two sectors (the first and second line in (\ref{R})) both being invariant under all the symmetries independently (up to ${\cal Z}_{AB} = 0$). This allows some freedom to fix the relative coefficient between both sectors, but we believe that this coefficient would be fixed as in (\ref{R}) due to supersymmetry, since it is the one required to match half-maximal supergravity in four dimensions \cite{Aldazabal:2011nj},\cite{Dibitetto:2012rk}.

~

On the other hand, generalized diffeomorphisms are generated by infinitesimal parameters $\xi^M = {\cal E}_A{}^M
\lambda^A$ in the fundamental representation of $G$ that take the form
\be \label{gaugeK}
\begin{split}
\delta_{\xi}
d =&  ~ \xi^{  M}\partial_{  M}   d -
\frac 12 \partial_{  M}   \xi^{  M}\  =\ \frac 1 2 \lambda^A {\cal F}_A - \frac 1 2 \cD_A \lambda^A \, ,   \\
\delta_{ \xi}  {\cal E}^A{}_M =& ~\xi^P\partial_P
 {\cal E}^A{}_M
                  +\left(\partial_{  M} \xi^{  P}-\partial^{  P}
 \xi_{  M} \right)  {\cal E}^A{}_P\ =\ {\cal E}_{BM} \left(2 \cD^{[B}\lambda^{A]} + {\cal F}^{AB}{}_C \lambda^C\right)\, .
     \end{split}
\ee
This further implies
\bea
 \delta_\xi {\cal F}_{ABC} &=& \lambda^D \cD_D {\cal F}_{ABC} + 4{\cal Z}_{ABCD}\lambda^D  +  3\cD_D \lambda_{[A} \Omega^D{}_{BC]}\, ,\label{f3var}\\
\delta_\xi {\cal F}_{A} &=& \lambda^D \cD_D {\cal F}_A + {\cal Z}_{AB}\lambda^B + {\cal F}^B \cD_B \lambda_A - \cD^B \cD_B \lambda_A + \Omega^C{}_{AB} \cD_C \lambda^B \, ,\label{f1var}
\eea
where
\be
    {\cal Z}_{ABCD} = \cD_{[A}{\cal F}_{BCD]}-\frac{3}{4}{\cal F}_{[AB}{}^E {\cal F}_{CD]E} = -\frac{3}{4}\Omega_{E[AB}\Omega^E{}_{CD]}\, ,\label{gf_ktens4}
\ee
and ${\cal Z}_{AB}$ was defined in (\ref{gf_ktens2}). Again, the failure of ${\cal F}_{ABC}$ and ${\cal F}_{A}$ to transform as scalars implies that DFT is a restricted theory and can only be consistently defined for a subset of fields and gauge parameters that ensure gauge invariance and closure. The quantity (\ref{gf_ktens4}) also vanishes if
(\ref{strong}) is imposed, but demanding that ${\cal F}_{ABC}$ and ${\cal F}_A$ transform as scalars only requires a relaxed version of the strong constraint
\bea
4{\cal Z}_{ABCD}\lambda^D  +  3\cD_D \lambda_{[A} \Omega^D{}_{BC]} &=&0 \ ,\nn\\
{\cal Z}_{AB}\lambda^B + {\cal F}^B \cD_B \lambda_A - \cD^B \cD_B \lambda_A + \Omega^C{}_{AB} \cD_C \lambda^B &=& 0 \ .\label{gaugeinvariance}
\eea
We will now show that both, invariance of the action under $H$-transformations (\ref{gaugeinv2}) and generalized diffeomorphisms (\ref{gaugeinvariance}) follow from closure.

Consider a gauge transformation for a generic tensorial density $V^M$ of
 weight $\omega(V)$
 \be
 \delta_{ \xi}  {V}^M = \xi^P\partial_P
 V^M                  +\left(\partial^{  M} \xi_{  P}-\partial_{  P}
 \xi^{  M} \right)  V^P + \omega(V) \partial_P \xi^P V^M\ ,
 \ee
the equations (\ref{gaugeK}) are then recovered for $\omega(e^{-2d}) = 1$ and $\omega({\cal E}) = 0$.
 These transformations  define the so-called C-bracket
\bea \left[ \xi_1, \xi_2\right]^{ M}_{\rm C} =\frac 12 \left(\delta_{\xi_1}\xi_2 - \delta_{\xi_2}\xi_1\right)^M&=&
2 \xi_{[1}^{ N}
\partial_{  N}
  \xi^{  M}_{2]} -   \xi^{  P}_{[1} \partial^{  M}
  \xi_{2]  P}  \nn\\ &=& {\cal E}_A{}^M \left(\left[\lambda_1, \lambda_2\right]^A_{\rm C} + {\cal F}_{BC}{}^{A} \lambda_1^B
\lambda_2^C\right) \, .\label{Cbracket} \eea
Generically, the commutator of two transformations of an arbitrary vector $V^M$  is not  a transformation, but differs as
  \be \left[  {\delta}_{ \xi_1},  {\delta}_{ \xi_2} \right] V^{ M} =
  {\delta}_{[ \xi_1, \xi_2]_{\rm C}} V^{ M}  - F^M(\xi_1, \xi_2, V) \, ,\label{DefC}\ee
where
\be
F^M (\xi_1, \xi_2, V) = \xi^Q_{[1} \partial^P \xi_{2]Q} \partial_P V^M + 2 \partial_P \xi_{[1 Q} \partial^P \xi_{2]}^M V^Q + \omega(\xi_3) \xi_{[1}^Q \partial_P\partial^P \xi_{2]Q} V^M  \label{FirstClosure}\ee
carries the same index structure as $V$. This indicates that the gauge transformation of a tensor is not automatically a tensor, and that the vanishing of its failure (denoted as $\Delta_{\xi}$) must be imposed as a constraint
\be
\Delta_{\xi_1} \delta_{\xi_2} V^M  = 0\ .\label{closuretotal}
\ee
The vanishing of $F^M$ in (\ref{DefC}) then follows from (\ref{closuretotal}). We will refer to (\ref{closuretotal}) as the {\it closure} constraints.
Notice that in particular  they imply
\bea
\Delta_{\xi_1} {\cal F}_{AB}{}^C &=& {\cal E}^C{}_M \Delta_{\xi_1} \delta_{{\cal E}_A} {\cal E}_B{}^M =  0\nn\\
\Delta_{\xi_1} {\cal F}_A &=& - e^{2d}\Delta_{\xi_1} \delta_{{\cal E}_A} e^{-2d} = 0
\eea
and then they guarantee that the dynamical fluxes transform as scalars under generalized diffeomorphisms, guaranteeing in turn the gauge invariance of the action, i.e. closure implies (\ref{gaugeinvariance}). Also, notice that due to closure
\be
{\cal Z}_{ABCD} = \Delta_{E_A} {\cal F}_{BCD} = 0  \ , \ \ \ \ \ \ {\cal Z}_{AB}  =\Delta_{E_A} {\cal F}_{B} = 0
\ee
and then $H$-invariance of the action (\ref{gaugeinv2}) is also guaranteed by closure.

Summarizing, closure requires the imposition of constraints (\ref{closuretotal}) that guarantee gauge invariance of the action, i.e. closure implies (\ref{gaugeinv2})  and (\ref{gaugeinvariance}). There are further constraints arising from their gauge transformed. Since they are known to admit solutions beyond the strong constraint \cite{Aldazabal:2011nj}, let us now briefly review the solutions of \cite{Aldazabal:2011nj} (which contain the strongly constrained case as a particular example). In the next section we will deal with geometry, and new constraints will arise,  which are also satisfied by these solutions.

\subsubsection{Scherk-Schwarz solutions}\label{SSsol}

All the constraints above are solved by restricting the fields as
\be
{\cal E}_A{}^M (X) = \widehat E_A{}^I (x) U_I{}^M (Y)\ , \ \ \ \ \ d = \widehat d (x) + \lambda (Y)\ ,
\ee
and the gauge parameters as
\be
\xi^M(X) = \lambda^A(x)\widehat E_A{}^I(x) U_I{}^M(Y) \ .\label{gaugeparamSS}
\ee
Here we have used the following notation for the coordinate dependence $X = (\tilde x, \tilde y; x, y)$, $Y = ( \tilde y , y)$. So, while the $Y$ coordinates are double and play the roll of internal coordinates in a SS compactification, the $x$ coordinates correspond to the un-doubled external space-time directions (the hats indicate dependence on $x$ only). For more details we refer to \cite{Aldazabal:2011nj}. This ansatz satisfies all the constraints, when $U(Y)$, which is an element of $O(D,D)$ called duality twist matrix, is {\it constrained} to satisfy
\begin{itemize}
\item $(U_I{}^M -\delta_I{}^M)\partial_M \widehat g = 0$
\item $f_{IJK} = 3 \tilde \Omega_{[IJK]} = const. \ , \ \ \ \ \ \tilde \Omega_{IJK} = U_I{}^M \partial_M U_J{}^N U_{KN} $
\item $f_{I} =  \tilde \Omega^J{}_{JI} + 2 U_I{}^M \partial_M \lambda = 0$
\item the quadratic constraints of half-maximal supergravity \cite{Schon:2006kz}
\be
f_{H[IJ} f^H{}_{KL]}=0 \ .\label{quadconstraints}
\ee
\end{itemize}
Moreover, the first, third and fourth conditions can be further relaxed through the introduction of a warp factor \cite{Aldazabal:2011nj} in order to account for gaugings in the fundamental representation of $O(D,D)$, but here we introduce this ansatz for simplicity. It was shown in \cite{Dibitetto:2012rk} that all the possible solutions to (\ref{quadconstraints}) can be reached by means of proper selections of duality twist matrices. Some solutions (the duality orbits of non-geometric fluxes) {\it require} truly double twist matrices, i.e. depending on both $y$ and $\tilde y$ in such a way that the strong constraint is violated, and no T-duality can be performed to get rid of the dual coordinate dependence.

Of course, there might be other solutions to these constraints, perhaps associated to other kind of compactifications. Let us emphasize that this ansatz {\it contains} the usual decompactified strong constrained case. In fact, taking $U = 1$, $\lambda = 0$ and the coordinates $x^i$ taking values $i= 1,\dots,D$, one obtains the usual situation analyzed in the literature. From the point of view of this ansatz, this is just a particular limit in which all the compact directions are decompactified.

For these configurations all the consistency constraints are satisfied. In fact, it can be checked that
\be
{\cal Z}_{ABCD} = 0 \ , \ \ \ \ \ {\cal Z}_{AB} = 0 \ ,\label{backgroundconstraints}
\ee
and also relations of the form
\be
\partial_M \lambda^A \partial^M \lambda^B = 0\ , \ \ \ \ \partial_M \partial^M \lambda^A =0\ , \ \ \ \ \Omega^D{}_{AB} {\cal D}_D \lambda^C = 0 \ ,
\ee
hold as well. Notice also that now the set of generalized diffeomorphisms has been reduced to a residual subgroup broken by the background. The SS ansatz can be thought of as a fixed background $U$, with perturbations $\widehat E$ around it, such that when this is plugged in the action and equations of motion one obtains an effective action for the perturbations. All these issues are discussed in \cite{Aldazabal:2011nj}, where the compactification to four-dimensions was shown to reproduce the electric sector of half-maximal gauged supergravity.

Under a SS reduction, the dynamical fluxes become
\bea
{\cal F}_{ABC} &=& \widehat F_{ABC} + f_{IJK} \widehat E_A{}^I\widehat E_B{}^J\widehat E_C{}^K \ , \ \ \ \ \ \ \ \ \widehat F_{ABC} = 3 \widehat \Omega_{[ABC]}\ ,\\
{\cal F}_A &=& \widehat \Omega^B{}_{BA} + 2 \widehat E_A{}^I \partial_I \widehat d\ ,
\eea
where
\be
\widehat \Omega_{ABC} = \widehat E_A{}^I \partial_I\widehat E_B{}^J\widehat E_{CJ}\ ,
\ee
so they are purely $x$-dependent, and all the truly double dependence has accommodated into the constant gaugings. This is in fact a generic feature of SS compactifications: covariant tensors with planar indices only depend on external coordinates.

We now continue without assuming this particular form of the fields and gauge parameters, but we will show that this ansatz also solves the forthcoming constraints in Section \ref{Geometry}.

\subsection{Equations of motion}\label{eoms}

The equations of motion of the DFT
action (\ref{action1}) (without the terms we denoted $\Delta_{SC}{\cal R}$) were derived and analyzed in \cite{Hull:2009mi} and \cite{EOMS}.  Here we obtain the equations of motion of the action (\ref{actionaction}).

The variations of the objects appearing in the flux formulation of DFT
with respect to $\cE_A{}^M$ and to $d$ are given by
\bea
    \d_\cE \Omega_{ABC} &=& \cD_A \D_{BC}
                        + \D_A{}^D\Omega_{DBC} + \D_B{}^D\Omega_{ADC} + \D_C{}^D\Omega_{ABD}\ ,\\
    \d_\cE \cF_{ABC}     &=& 3\left(\cD_{[A} \D_{BC]}+ \D_{[A}{}^D\cF_{BC]D}\right)\ ,\\
    \d_\cE \cF_{A}       &=& \cD^B \D_{BA}+ \D_{A}{}^B\cF_{B}\ ,\\
    \d_d \cF_{A}         &=& 2\cD_A\,\d d \ ,
\eea
and these in turn translate into variations of the action (\ref{actionaction})  given by
\bea
    \d_\cE S &=& \int d  X e^{-2d}\; \cG^{AB} \D_{AB}\ ,\\
    \d_d S &=& \int d  X e^{-2d}\; \cG \d d\ ,
\eea
where
\be
    \D_{AB} = \d \cE_A{}^M \cE_{BM}= - \D_{BA}\ ,\label{antisym}
\ee
must be antisymmetric to enforce the constraint $\cE_A{}^M\cE_{BM} = \eta_{AB}$.
The equations of motion are then
\bea
    \cG^{[AB]} &=& 0\ , \label{eoms2}\\
    \cG &=&0\  ,
\eea
where
\bea
    \cG^{[AB]} 	&=& 2 (S^{D[A} - \eta^{D[A}) {\cal D}^{B]} {\cal F}_D +  ({\cal F}_ D - {\cal D}_D) \check{{\cal F}}^{D[AB]} +  \check{{\cal F}}^{CD[A}{\cal F}_{CD}{}^{B]}\\
    &=& {\cal Z}^{AB} + 2  S^{D[A} {\cal D}^{B]} {\cal F}_D +  ({\cal F}_ D - {\cal D}_D) \breve{{\cal F}}^{D[AB]} +  \breve{{\cal F}}^{CD[A}{\cal F}_{CD}{}^{B]}\, ,\nn\\
    \cG  &=& -2\cR\, .
\eea
Here, we have introduced the notation
\be
\check{{\cal F}}^{ABC}
= \check{S}^{ABCDEF}\ {\cal F}_{DEF} \, ,\qquad
\breve{{\cal F}}^{ABC} = \check{{\cal F}}^{ABC}  +
 {\cal F}^{ABC}\  ,
\ee
where
\bea
\check{ S}^{ABCDEF} &=& \frac 1 2 S^{AD}\eta^{BE}\eta^{CF}
                    +  \frac 1 2  \eta^{AD}S^{BE}\eta^{CF}
                    +  \frac 1 2  \eta^{AD}\eta^{BE}S^{CF}
                    -  \frac 1 2  S^{AD}S^{BE}S^{CF}\nn\\&& -  \eta^{AD}\eta^{BE} \eta^{CF} \nn\\
                    &=& \breve{S}^{ABCDEF} -  \eta^{AD}\eta^{BE} \eta^{CF} \ .\label{cupdefinitions}
\eea
The  operator $\breve{S}$ defines an involutive map $\breve{S}^2 = 1$, so $-  \check{S} / 2$ is a projector.

~

In the next section, these equations of motion will be re-obtained from a generalized notion of Ricci flatness.
%%%%%%%%%%%%%%%%%%%%%%%%%%%%%%%%%%%%%%%%%%%%%%%%%%%%%%%%%%%%%%%%%%%%%%%%%%%%%%%%%%%%%%%%%%%%%%%

\section{Geometry, connections and curvature}\label{Geometry}

It was shown in \cite{Siegel:1993th},\cite{DFTgeomHohmKwak},\cite{Geom}$-$\cite{WaldramR} that the action and equations of motion of DFT can be obtained from traces and projections of a generalized Riemann tensor. The construction goes beyond Riemannian geometry because it is based on the generalized  rather than the standard Lie derivative. Then, the notions of connections, torsion and curvature have to be generalized and many interesting features arise in this framework. For example, it turns out that
the   vanishing torsion and compatibility conditions do not completely  determine the
connections and curvatures but only fix some of their projections.
The strong constraint was always assumed in these constructions. In this section
we re-examine these generalized objects without imposing the strong constraint, but only the relaxed
constraints discussed in the previous section, plus new ones arising here. Our route will closely follow that of \cite{Geom}.

%%%%%%%%%%%%%%%%%%%%%%%%%%%%%%%%%%%%%%%%%%%%%%%%%%%%%%%%%%%%%%%%%%%%%%%%%%%%%%%%%%%%%%%%%%%%%%%
\subsection{Generalized connections}

We begin by defining a covariant derivative acting on tensors with curved and/or planar indices as
\begin{equation}
\nabla_M V_A{}^K =\partial_M V_A{}^K + \Gamma_{M N}{}^K V_A{}^N - \omega_{M A}{}^BV_B{}^K\ ,
\end{equation}
where $\Gamma_{MN}{}^K$ is a Christoffel connection,  and $\omega_{M A}{}^{B}$ a spin connection. The forthcoming list of conditions
were imposed in \cite{Siegel:1993th},\cite{DFTgeomHohmKwak},\cite{Geom}$-$\cite{WaldramR} to restrict these connections, following a
similar procedure to the usual one in Riemannian geometry.  The list is ordered in such a way that each item assumes the previous ones.

\begin{itemize}
%%%%%%%%%%%%%%%%%%%%%%%%%%%%%%%%%%%%%%%%%%%%%%%%%%%%%%%%%%%%%%%%%%%%%%%

%%%%%%%%%%%%%%%%%%%%%%%%%%%%%%%%%%%%%%%%%%%%%%%%%%%%%%%%%%%%%%%%%%%%%%%%%%%
\item {\bf Compatibility with the generalized frame}. Covariant constancy of $\mathcal{E}_A{}^N$
\be
\nabla_M {\cal E}_A{}^N = 0\ ,
\ee
relates the Christoffel, spin and Weitzenb\"ock connections
\begin{equation}\label{ThirdCondition}
\Gamma_{ML}{}^N= -\Omega_{ML}{}^N + \mathcal{E}^A{}_L\mathcal{E}_B{}^N\omega_{MA}{}^B\ .
\end{equation}
Since the Weitzenb\"ock connection is fully determined by the generalized frame, this condition simply relates the Christoffel and spin connections.
%%%%%%%%%%%%%%%%%%%%%%%%%%%%%%%%%%%%%%%%%%%%%%%%%%%%%%%%%%%

\item {\bf Compatibility with the $O(D,D)$ invariant metric}.
Given the covariant constancy of the generalized frame, covariant constancy of the metric $\eta^{MN}$
can be equally cast as
\begin{equation}
\nabla_{M}\eta^{NP}=0 ~~~\Longleftrightarrow ~~~ \nabla_{M} \eta^{AB}=0\, ,
\end{equation}
which in turn imply
\begin{equation}\label{FourthCondition}
\Gamma_{MNP}=- \Gamma_{MPN}~~~\Longleftrightarrow ~~~ \omega_{MAB}=-\omega_{MBA}\ .
\end{equation}

%%%%%%%%%%%%%%%%%%%%%%%%%%%%%%%%%%%%%%%%%%%%%%%%%%%%%%%%%

\item {\bf Compatibility with the generalized metric}. Covariant constancy of the generalized metric
\begin{equation}
\nabla_{M}\mathcal{H}_{NK}=0 ~~~ \Longleftrightarrow ~~~ \nabla_M S_{AB}=0\, ,
\end{equation}
implies that
\begin{equation}\label{FifthCondition}
\partial_M \mathcal{H}_{NK} - \Gamma_{MN}{}^P \mathcal{H}_{PK} - \Gamma_{MK}{}^P\mathcal{H}_{NP}=0 ~ \Longleftrightarrow ~ \omega_{MA\check{B}} = - \omega_{MB\check{A}}\ .
\end{equation}
Here we used the check notation for indices contracted with the planar generalized metric (\ref{Check}).
%%%%%%%%%%%%%%%%%%%%%%%%%%%%%%%%%%%%%%%%%%%%%%%%%%%%%%%%%%%%%%%%%%%%%%%%%

\item {\bf Covariance under generalized diffeomorphisms}.
The spin connection is requested to transform covariantly under generalized diffeomorphisms
\be
\delta_\xi \omega_{AB}{}^C = \xi^P \partial_P  \omega_{AB}{}^C\, .
\ee
 Through bein compatibility we then have
\begin{equation}\label{eq:varcovderSC}
\Delta_\xi \Gamma_{MNP}  = - \Delta_\xi \Omega_{MNP} = 2 \partial_M \partial_{[N}\xi_{P]} - \partial_Q \xi_M \Omega^Q{}_{NP}\, ,
\end{equation}
where we define $\Delta_\xi$ as the failure of an expression to transform covariantly.

%%%%%%%%%%%%%%%%%%%%%%%%%%%%%%%%%%%%%%%%%%%%%%%%%%%%%%%%%%%%%%%%%%%%%%%%%%%%%%

\item {\bf Covariance under double Lorentz transformations}. Under local H transformations, we demand that $\nabla_M V_A{}^K$ transforms as a Lorentz vector. This implies that
\begin{equation}
\delta_\Lambda \Gamma_{MN}{}^K=0\ ,
\end{equation}
and
\begin{equation}\label{variationlorentzomega}
\delta_\Lambda \omega_{MA}{}^B= \partial_M\Lambda_A{}^B + \omega_{MC}{}^B\Lambda_A{}^C-\omega_{MA}{}^C\Lambda_C{}^B\ .
\end{equation}

\item {\bf Vanishing generalized torsion}. The standard definition of torsion turns out to be non-covariant under generalized diffeomorphisms. Then, one has to resort to a generalized definition \cite{WaldramR}
\begin{equation}
(\delta_{\xi}^\nabla - \delta_{\xi})V^M =  \mathcal{T}_{QP}{}^M \xi^Q V^P,
\end{equation}
where $V^M$ is a vector and  $\delta^\nabla$ is the generalized gauge transformation with $\partial_M$ replaced by $\nabla_M$. This definition yields
\begin{equation}
\mathcal{T}_{QP}{}^M = 2 \Gamma_{[QP]}{}^M  - \Gamma^M{}_{PQ}\, .
\end{equation}

Combined with compatibility with the $O(D,D)$ metric, one finds that
\be \mathcal{T}_{MNK}=3\Gamma_{[MNK]}  ~~ \Longleftrightarrow ~~ {\cal T}_{ABC} = 3 \omega_{[ABC]} - {\cal F}_{ABC}  \, ,\label{gentorsion}\ee
 and then setting the torsion to zero, we obtain
\begin{equation}\label{SixthCondition}
\Gamma_{[MNK]}=0 ~~ \Longleftrightarrow ~~ \mathcal{F}_{ABC}=3 \omega_{[ABC]}\, .
\end{equation}
Note that this condition is consistent with the transformation properties of $\mathcal{F}_{ABC}$  under generalized diffeomorphisms provided the gauge consistency constraints hold. The antisymmetrization of the spin connection (which is requested to be covariant) coincides with the dynamical fluxes, which were also requested to be covariant. It then follows from the constraints that  the generalized torsion is covariant as well.
%%%%%%%%%%%%%%%%%%%%%%%%%%%%%%%%%%%%%%%%%%%%%%%%%%%%%%%%%%%%%%%%%%%

\item {\bf Compatibility with the generalized dilaton}. Demanding partial integration in the presence of the dilaton measure $e^{-2d}$:
\be
\int e^{-2d} W \nabla_M U^M = - \int e^{-2d} U^M \nabla_M W\, ,
\ee
one finds
\begin{equation}\label{SeventhCondition}
\Gamma_{PM}{}^P=-2\partial_Md ~~ \Longleftrightarrow ~~ \omega^B{}_{BA}=\mathcal{F}_A\ .
\end{equation}
Again we find consistency in requiring that the spin connection is covariant, because its trace is related to the dynamical fluxes which are covariant as well.

\end{itemize}
It was shown in \cite{WaldramR}$-$\cite{Geom} that these constraints only determine some projections of the connections, leaving undetermined pieces which cannot be identified with the physical degrees of freedom.  Still, some projections of a generalized Riemann tensor reproduce the action and equations of motion. In some cases \cite{Park} some further projections on the connection are requested to vanish in order to eliminate the undetermined part. However, in this case
the derivative is only covariant under particular projections and then full covariance is lost.
More recently, in \cite{BermanGEOM} the connection was chosen to equal the Weitzenb\"ock connection, and then the spin connection vanishes. The advantage of the construction in \cite{BermanGEOM} is that the connection is simple and determined. The torsion (\ref{gentorsion}) is non-vanishing and equals the antisymmetric part of the Weitzenb\"ock connection, so it coincides with the dynamical fluxes discussed here. Although the (generalized) connection is
 flat, the dynamics is encoded in the torsion, and the action is constructed by demanding $H$-invariance. Interestingly, the strong constraint can be relaxed in this formulation as well.
Here, we will follow the route of \cite{WaldramR}$-$\cite{Geom}, obtaining the action and equations of motion from traces of the generalized Riemann tensor. We will show that this provides a systematic way of obtaining the full action (\ref{actionaction}), and equations of motion (\ref{eoms2}).

Imposing the additional constraint that the spin connection is linear in fluxes,  a unique solution to (\ref{FourthCondition}), (\ref{FifthCondition}), (\ref{SixthCondition}), (\ref{SeventhCondition}) can be found
\begin{equation}\label{solutionlinearorder}
\begin{split}
\omega_{ABC} =& -\frac{1}{D-1} \left( \mathcal{F}_{[B}\eta_{C]A} + \mathcal{F}^D S_{D[B}S_{C]A} \right)\\
& + \frac{1}{3}\left( \mathcal{F}_{ABC} + \mathcal{F}_{A\check{B}\check{C}} - \frac{1}{2}\mathcal{F}_{\check{A}\check{B}C} -\frac{1}{2} \mathcal{F}_{\check{A}B\check{C}} \right) \, .
\end{split}
\end{equation}
However,  a covariant derivative built from this particular connection does not satisfy (\ref{variationlorentzomega}) under $H$-transformations, only some projections do, and then this connection is semi-covariant. In what follows we will not make use of (\ref{solutionlinearorder}), but instead we will work only with the previous conditions on the connections.

 Notice that due to the above requirements, the derivative of the spin connection is required to transform as a tensor under generalized diffeomorphisms
\be
\Delta_\xi \partial_M \omega_{AB}{}^C = \partial^P \xi_M \partial_P \omega_{AB}{}^C = 0 \, .\label{newconstr2}
\ee
Moreover, due to (\ref{eq:varcovderSC}) we have an additional constraint from covariance of the covariant derivative
\be
\Delta_\xi \nabla_M V_N = \Delta_\xi \left[\partial_M V_N - \Gamma_{MNP} V^P\right] = 0\, ,
\ee
which can be recast in the form
\be
\partial_P \xi_M \partial^P V_N + \partial_P \xi_M \Omega^P{}_{NQ} V^Q =0 \, .\label{newconstr3}
\ee
We now have new constraints, for the vectors, gauge parameters and connections, like (\ref{newconstr2}) and (\ref{newconstr3}), that arise by demanding that this geometric construction is consistent with a relaxation of the strong constraint. Notice that these constraints are not requested for consistency of the theory. Moreover, only some projections of them are physical, because of the undetermined components of the connection. In any case, as strong as they look, they are all  satisfied once again by the SS solutions of Section  \ref{SSsol}. In fact, as we explained in that section, in the SS scenario the covariant objects in planar indices only depend on the external coordinates, and then it is easy to see that (\ref{newconstr2}) is satisfied in a SS reduction where the gauge parameters take the form (\ref{gaugeparamSS}). This is consistent with the fact that projections of the spin connections give generalized fluxes, which also only depend on the external coordinates in this case. As for (\ref{newconstr3}), notice that the strong constraint terms of the form $\Omega^Q{}_{MN} \Omega_{QRS}$ cancel, so it is also satisfied by the SS ansatz. Then, these new constraints are also solved by truly double SS reductions, but more generally might be solved by other truly double configurations.
%%%%%%%%%%%%%%%%%%%%%%%%%%%%%%%%%%%%%%%%%%%%%%%%%%%%%%%%%%%%%%%%%%%%%%%%%%%%%%%%%%%%%%%%%%%%%%%

\subsection{Generalized curvature}\label{curvatures}

The usual Riemann tensor in planar indices (i.e., rotated with the bein)
\be
	R_{ABC}{}^D =	2 \left(\cD_{[A}\omega_{B]C}{}^D - \Omega_{[AB]}{}^E \omega_{EC}{}^D -
		 \omega_{[A\vert C}{}^E\omega_{\vert B]E}{}^D\right)\ ,
\ee
is not a scalar under generalized diffeomorphisms (even if the strong constraint were imposed) because the Weitzenb\"ock connection is not covariant. However, following the steps of \cite{Siegel:1993th},\cite{DFTgeomHohmKwak},\cite{WaldramR}-\cite{Geom} one can extend this definition in order to covariantize it\footnote{Imposing
the vanishing of its failure to transform covariantly  as a new constraint, is not an option. We are assuming that all the constraints of DFT are solved by the strong constraint, so that there is always a limit that makes contact with supergravity. }. Consider for example the following modified curvature
\be\begin{split}
	\hat{R}_{ABCD} &=
	R_{ABCD} - \Omega^E{}_{AB}\omega_{ECD} \\
	&= 2\cD_{[A}\omega_{B]CD} - \cF_{AB}{}^E \omega_{ECD}
		- 2\omega_{[A\vert C}{}^E\omega_{\vert B]ED}\ . \label{Rhat}
\end{split}\ee
An extra term is included in order to promote the Weitzenb\"ock connection to a generalized flux, which is covariant. This expression is now a scalar under generalized diffeomorphisms. With the addition of the new term in (\ref{Rhat}), the $G_L$ covariance has now been compromised. In order to restore it we  further extend the definition as \cite{Siegel:1993th}
\be\label{eq:RcalABCD}
\begin{split}
	\cR_{ABCD} &=
	 \hat{R}_{ABCD} + \hat{R}_{CDAB} + \omega^E{}_{AB}\,\omega_{ECD}\ , \\
	\end{split}\ee
which is also a scalar under generalized diffeomorphisms. Of course, we are expecting that $G_L$ or $H$ invariance is achieved only up to strong constraint violating terms, because so is the action (\ref{VarHaction}). A quick computation shows that
\be
\Delta_\Lambda {\cal R}_{ABCD} = {\cal D}_E \Lambda_{BA} \Omega^E{}_{CD} + {\cal D}_E \Lambda_{DC} \Omega^E{}_{AB}\ ,
\ee
so if one pretends a fully covariant Riemann tensor, this must be set to zero. In particular, under a SS reduction  $\Lambda_{AB}$ would depend on external coordinates only, and this constraint would be automatically satisfied.

Rotating all indices with the generalized bein, and using (\ref{ThirdCondition}), the generalized Riemann tensor in curved indices can be cast in the form
\begin{equation}\label{eq:RcalMNKL}
\mathcal{R}_{MNKL}= \hat{\mathcal{R}}_{MNKL} - \Omega_{QMN}\Omega^Q{}_{KL}\ ,
\end{equation}
where
\begin{equation}\label{eq:hatRcalMNKL}
\begin{split}
& \hat{\mathcal{R}}_{MNKL}=R_{MNKL} + R_{KLMN} + \Gamma_{QMN}\Gamma^{Q}{}_{KL}\ ,\\
& R_{MNKL}=2\partial_{[M}\Gamma_{N]KL} + 2\Gamma_{[M|QL}\Gamma_{|N]K}{}^{Q}\ .
\end{split}
\end{equation}
Here, $\hat{\mathcal{R}}_{MNKL}$ is the generalized Riemann tensor found in \cite{Geom}. We see that the difference between (\ref{eq:RcalMNKL}) and (\ref{eq:hatRcalMNKL}) is a strong constraint-violating term which does not vanish with our assumptions. This extra factor was also considered in \cite{Aldazabal:2013mya}, where the first geometric construction with a relaxed strong constraint was built in the U-duality case. The generalized Riemann tensor (\ref{eq:RcalMNKL}) enjoys the same symmetry properties of the usual one, namely ${\cal R}_{MNKL} = {\cal R}_{([MN][KL])}$.

Following the path of \cite{Geom} we now want to consider traces and projections of the generalized Riemann tensor to get a generalized Ricci tensor and scalar. For instance, imposing (\ref{SixthCondition}) and (\ref{SeventhCondition}), we obtain
\be\begin{split}\label{curvbi}
	\cR_{AB}{}^{AB} &=  -4\cZ\ ,\\	
\end{split}\ee
where ${\cal Z}$ was defined in (\ref{gf_ktens0}). This vanishes under the strong constraint, but here it gives rise to some of the strong constraint-violating terms in the action. On the other hand, contractions with $S$ (or $\mathcal{H}$) give the same answer
\begin{equation}\label{eq:curvibSS}
\mathcal{R}_{\check{A}\check{B}}{}^{AB}= -4\mathcal{Z}\ .
\end{equation}
 Thus, we are led to consider traces of the generalized Riemann tensor with mixed $S^{AC}$ and $\eta^{BD}$ contractions. After imposing conditions (\ref{FourthCondition}), (\ref{FifthCondition}), (\ref{SixthCondition}), (\ref{SeventhCondition}), all the undetermined parts of the connection drop out from (\ref{eq:RcalABCD}) and one gets
\be\label{curvbic}
	\cR_{\check{A}B}{}^{AB} = -2 \cR -4 {\cal Z}\ .
\ee

In order to combine these results we introduce the projectors
\begin{equation}
\begin{split}
&P_{M}{}^{N}=\frac{1}{2}\left(\delta_{M}{}^N - \mathcal{H}_M{}^N \right)~~\mbox{or}~~P_{A}{}^B=\mathcal{E}_A{}^M\mathcal{E}^B{}_{N}P_M{}^N=\frac{1}{2}\left(\delta_A{}^B - S_A{}^B \right),\\
&\bar{P}_{M}{}^{N}=\frac{1}{2}\left(\delta_{M}{}^N + \mathcal{H}_M{}^N \right)~~\mbox{or}~~\bar{P}_{A}{}^B=\mathcal{E}_A{}^M\mathcal{E}^B{}_{N}\bar{P}_M{}^N=\frac{1}{2}\left(\delta_A{}^B + S_A{}^B \right).
\end{split}
\end{equation}
Using the results (\ref{curvbi}), (\ref{eq:curvibSS}) and (\ref{curvbic}) we see that the unique combination giving the full generalized Ricci scalar  in terms of projectors is
\begin{equation}
\mathcal{R}  = \frac{1}{4}P^{AC}P^{BD}\mathcal{R}_{ABCD}\  ,
\end{equation}
where $\mathcal{R}$ was defined in (\ref{R}).\footnote{
Other combinations give
\begin{equation}
\bar{P}^{MK}P^{NL}\mathcal{R}_{MNKL}=P^{MK}\bar{P}^{NL}\mathcal{R}_{MNKL}=0,
\end{equation}
\begin{equation}
\bar{P}^{MK}\bar{P}^{NL}\mathcal{R}_{MNKL}=-4\mathcal{R} - 16\mathcal{Z} .
\end{equation}
Note the difference between acting with $PP$ and $\bar{P}\bar{P}$ on $\mathcal{R}_{MNKL}$ when the strong constraint is relaxed.}

 Also, the completely antisymmetric part of $\cR_{ABCD}$ only involves the antisymmetric parts of the connection. Imposing
(\ref{SixthCondition}) and (\ref{SeventhCondition}) again, we obtain from (\ref{eq:RcalABCD}) an algebraic BI for the generalized Riemann tensor
\begin{equation}\label{curvbiantisym}
\cR_{[ABCD]} = \frac{4}{3}\cD_{[A}\cF_{BCD]} - \cF_{[AB}{}^E\cF_{CD]E}
	=  \frac{4}{3}\cZ_{ABCD}\, .
\end{equation}
Identities like this, and many others are extensively discussed in \cite{Geom}.

%%%%%%%%%%%%%%%%%%%%%%%%%%%%%%%%%%%%%%%%%%%%%%%%%%%%%%%%%%%%%%%%%%%%%%%%%%%%%%%%%

%%%%%%%%%%%%%%%%%%%%%%%%%%%%%%%%%%%%%%%%%%%%%%%%%%%%%%%%%%%%%%%%%%%%%%%%%%%%%%%%%%%%%%%%%%%
\subsection{Generalized Ricci flatness}
The full action (\ref{actionaction}) can be written as
\be\label{eq:fullactionconN=4terms}
	S  = \frac{1}{4}\int dX\ e^{-2d}\ P^{MK}P^{NL}\mathcal{R}_{MNKL}\ ,
\ee
and its variation with respect to the bein $\mathcal{E}$ gives
\begin{equation}\label{eq:varepsilonfullS}
\delta_{\mathcal{E}}S= \frac{1}{4}\int d X\ e^{-2d}\ \left( 2(\delta_{\mathcal{E}}P^{MK})P^{NL}\mathcal{R}_{MNKL} + P^{MK}P^{NL}\delta_{\mathcal{E}}\mathcal{R}_{MNKL}\right).
\end{equation}
The projectors satisfy $P^2=P$, $\bar{P}^2=\bar{P}$, $P + \bar P = 1$ and $P \bar P = 0$, and we require that the shifted ones
$P'=P+\delta_{\mathcal{E}} P$ (or $\bar{P}'$)
also obey these relations. This implies that
\begin{equation}\label{eq:varepsilondeP}
\delta_{\mathcal{E}}P^{MK}= P^M{}_R\delta_{\mathcal{E}}P^{RL}\bar{P}^K{}_L + \bar{P}^M{}_L\delta_{\mathcal{E}}P^{LR}P_{R}{}^K\ .
\end{equation}
Also, by  definition we have
\be
\delta_{\mathcal{E}}P^{RL}=-\frac{1}{2}\left( \delta\mathcal{E}_A{}^RS^{AB}\mathcal{E}_{B}{}^L + \mathcal{E}_A{}^RS^{AB}\delta\mathcal{E}_{B}{}^L \right)\ ,
 \ee and inserting this information in the first term of (\ref{eq:varepsilondeP}) we find
\begin{equation}
2(\delta_{\mathcal{E}}P^{MK})P^{NL}\mathcal{R}_{MNKL} =-4\Delta_{AC}\ P^{BC}\bar{P}^{DA}P^{EF}\mathcal{R}_{BEDF}\, ,
\end{equation}
where we used (\ref{antisym}). Recalling (\ref{eq:RcalMNKL}),  the second term of (\ref{eq:varepsilonfullS}) is
\begin{equation}\label{eq:PPvarRhat}
\int dX\ e^{-2d}\ P^{MK}P^{NL}\delta_{\mathcal{E}}\mathcal{R}_{MNKL}= \int dX\ e^{-2d}\ P^{MK}P^{NL}\delta_{\mathcal{E}}(\hat{\mathcal{R}}_{MNKL}-\Omega_{QMN}\Omega^{Q}{}_{KL})\ .
\end{equation}

The infinitesimal variation of $\hat{\mathcal{R}}_{MNKL}$ with respect to $\mathcal{E}$ can be computed by first varying with respect to $\Gamma$ \cite{Geom}
\begin{equation}
\delta_{\mathcal{E}}\hat{\mathcal{R}}_{MNKL}=2\nabla_{[M}\delta_{\mathcal{E}}\Gamma_{N]KL} + 2\nabla_{[K}\delta_{\mathcal{E}}\Gamma_{L]MN}\ .
\end{equation}
Inserting this variation into (\ref{eq:PPvarRhat}), the projectors pass through the covariant derivative (since $\nabla\eta=\nabla\mathcal{H}=0$) and we get a total derivative, due to the dilaton compatibility condition. The second term of (\ref{eq:PPvarRhat}) gives
\be
\int dX\ e^{-2d}\ P^{MK}P^{NL}\delta_{\mathcal{E}}(\Omega_{QMN}\Omega^{Q}{}_{KL})
=-2\int dX\ e^{-2d}\ \Delta_{AC}P^{AE}P^{CF} \mathcal{Z}_{EF}\ .
\ee
Putting all this together, we finally get
\begin{equation}
\delta_{\mathcal{E}}S=\frac 1 4\int dX\ e^{-2d}\ \Delta_{AC} P^{BC}\bar{P}^{AD} ( - 4 P^{EF}\mathcal{R}_{BEDF} - 2 {\cal Z}_{BD}) = \int dX\ e^{-2d}\ \Delta_{AC}\ {\cal G}^{[AC]}\, .
\end{equation}
Then the equations of motion are
\begin{equation}
{\cal G}^{[AC]} =   P^{B[A}\bar{P}^{C]D} \left( P^{EF}\mathcal{R}_{BEDF} + \frac 1 2 {\cal Z}_{BD}\right) =0\, ,
\end{equation}
which match  those found in (\ref{eoms2}).

It might seem surprising at first sight that this form of generalized Ricci flatness is governed by an antisymmetric tensor. We recall however that there is a remarkable property of the  projections with $P$ and $\bar P$
\be \begin{matrix}
{P}_{M}{}^R  \bar{P}_{N}{}^S K_{RS}= 0&\Rightarrow &  {P}_{[M}{}^R  \bar{P}_{N]}{}^S K_{RS} = 0& \Rightarrow &   P_{Q}{}^M  {P}_{[M}{}^R  \bar{P}_{N]}{}^S K_{RS} = 0\\
\Uparrow & &\Updownarrow & & \Downarrow\\
  P_Q{}^M {P}_{(M}{}^R  \bar{P}_{N)}{}^S K_{RS}= 0& \Leftarrow&  {P}_{(M}{}^R  \bar{P}_{N)}{}^S K_{RS} = 0& \Leftarrow &  {P}_{M}{}^R  \bar{P}_{N}{}^S K_{RS} = 0 \end{matrix}
\ee
Namely, the symmetric and antisymmetric pieces contain the same information. Then, it is possible to define a {\it symmetric} generalized Ricci tensor, whose flatness gives the equations of motion as well
\be
{\cal R}^{AC} =  P^{B(A}\bar{P}^{C)D} \left(P^{EF}\mathcal{R}_{BEDF} + \frac 1 2 {\cal Z}_{BD}\right) =0\ .
\ee
%%%%%%%%%%%%%%%%%%%%%%%%%%%%%%%%%%%%%%%%%%%%%%%%%%%%%%%%%%%%%%%%%%%%%%%%%%%%%%%%%%%%%%%%%%%%%%%

%%%%%%%%%%%%%%%%%%%%%%%%%%%%%%%%%%%%%%%%%%%%%%%%%%%%%%%%%%%%%%%%%%%%%%%%%%%%%%%%%%%%%%%%%%%%%%%
\section{Type II and Heterotic DFT}\label{TypeIIhet}

\subsection{Type II }

In addition to the NS-NS sector, type II supergravity has a set of p-form gauge fields, $C_1$ and $C_3$ for type IIA or $C_0$, $C_2$ and $C_4$ for IIB, belonging to the R-R sector. The inclusion of R-R fields was extensively addressed in \cite{TypeII},\cite{WaldramR}. Here we only intend to relate the constraints in this sector with the results of the previous sections. In the so-called democratic formulation, the set of gauge field strengths $G_p$ is completed by magnetic duals $G_{10-p}$ and packed in a sum of differential forms, or polyform,
\be\label{df_poly}
    {\mathbb G} = \sum_{p=0,1}^{10,9} G_p
    = \sum_{p=0,1}^{10,9} \frac{1}{p!}G_{i_1...i_p} dx^{i_1}\wedge...\wedge dx^{i_p}\ ,
\ee
where $p$ is odd for IIB or even for IIA. To recover the correct number of degrees of freedom, a self-duality condition is imposed by hand on the total field strength
\be\label{df_sd}
    {\mathbb G} = \star \sigma {\mathbb G}\ ,
\ee
where $\star$ is the Hodge star and $\sigma$ is an involution reversing the order of the differentials $dx^i$, or equivalently flipping the sign for $p =2,3 \mod 4$. The total field strength ${\mathbb G}$ descends from a gauge potential polyform
${\mathbb C}= C_{0,1} + C_{2,3} + \dots$ which contains all the electric and magnetic potentials \cite{oddrr}
\be\label{df_rfs}
\begin{split}
    {\mathbb G} &= (d+H\wedge){\mathbb C} + me^{-B}\ ,\\
    (d+H\wedge){\mathbb G} &=0\ ,
\end{split}
\ee
where $m = G_0$ is Roman's mass parameter. Notice that the twisted exterior derivative $d+H\wedge$ is nilpotent due to the BI of the NS-NS three-form $dH=0$. The field strengths $H$ and ${\mathbb G}$ are invariant under the following gauge transformations
\be\label{df_gauge}
\begin{split}
    \d B &= d\lambda\ ,\\
    \d {\mathbb C} &= (d+H\wedge)\L +m\,\lambda\wedge e^{-B}\ ,
\end{split}
\ee
respectively,
where $\lambda$ is an arbitrary one-form and $\L = \L_{0,1} + \L_{2,3} + ... $ is an arbitrary polyform.

The total field strength ${\mathbb G}$ transforms as an $O(10,10)$ spinor under T-duality. Since $D$-dimensional polyforms live in a spinorial representation of $G=O(D,D)$, it is natural to consider the R-R fields as $O(D,D)$ spinors in DFT, as achieved in \cite{TypeII}. When the theory is formulated in terms of the $G$-singlets $\cF_{ABC}$ and $\cF_A$, a possible formulation is to take R-R fields in a representation of $G_L=O(D,D)$ while keeping them invariant under $G$. Roman's mass $m$ will be set to zero in what follows, we refer the reader to \cite{Massive} for a DFT treatment with a non-vanishing value.

For the signature $(D,D)$, there always exist real gamma matrices $\G^A = (\G^a, \G_a)$ giving a representation of the $G_L=O(D,D)$ Clifford algebra $\{\G^A,\G^B\} = \eta^{AB}$. Since the matrices $(\G^a, \G_a)$ span a fermonic oscillator algebra $\{\G^a,\G_b\} = \d^a_b$, any polyform such as ${\mathbb G}$ can be mapped to an $O(D,D)$ spinor $\cG$ as
\be\label{rr_exp}
    \cG = \sum_p \frac{e^\phi}{p!}
		G_{i_1...i_p}\, e_{a_1}{}^{i_1}...e_{a_p}{}^{i_p}\, \G^{a_1 ... a_p}\vert 0\rangle\ ,
\ee
where $\vert 0\rangle$ is a Clifford vacuum annihilated by $\G_a$ and where the dilaton factor has been added for convenience. For $O(D,D)$ spinors it is possible to find a matrix that mimics the effect of the operator $\star\sigma$ when acting on a spinor written as in (\ref{rr_exp}). For $D=1+9$ this operator reads
\be\label{rr_hplus}
    \Psi_+ =(\G^0 - \G_0)(\G^1 + \G_1) ...(\G^9 +\G_9)\ ,
\ee
and squares to the identity. We refer the reader to appendix \ref{oddspin} for the definition and properties of this operator for generic dimension. The self-duality condition (\ref{df_sd}) can then be implemented on the spinor $\cG$ by
\be\label{rr_dual}
    \cG = \Psi_+ \cG\ .
\ee
We also note that the (anti-)chirality condition on the spinor $\cG$ is simply translated by (odd) even forms in the expansion (\ref{rr_exp}), so that the spinorial field strength is chiral for IIB and anti-chiral for IIA. Being a spinor, the field strength $\cG$ transforms under $G_L$ as
\be\label{rr_llt}
    \d \cG = \frac{1}{2}\L_{AB}\G^{AB}\cG\ .
\ee
It is then possible to build a derivative operator $\nabla_A$, in a way that $\nabla_A\cG$ transforms covariantly under $G_L$.

 ~

  When (\ref{FourthCondition}) is satisfied, the covariant derivative can be extended to act in any representation of $G_L$, with generators $\Sigma^{AB}$ and Lorentz algebra $[\Sigma^{AB}, \Sigma^{CD}]=4\eta^{[A|[C}\Sigma^{D]|B]}$. In order to have only explicit Lorentz indices, a covariant derivative and connection with flat indices can be defined
\begin{equation}
\nabla_AT=\mathcal{E}_A{}^M \nabla_M T=\left( \mathcal{D}_A -\frac{1}{2}\omega_{ABC}\Sigma^{BC} \right)T\ ,
\end{equation}
where $T$ generically transforms as $\delta_\Lambda T=\frac{1}{2}\Lambda_{AB}\Sigma^{AB}T$ and it is a scalar under generalized diffeomorphisms. For $\nabla_A T$ to transform as a scalar under generalized diffeomorphisms parameterized,  $\omega_{ABC}$ shall transform as a scalar and the following constraint
\be
	\partial^M \xi^N \, \partial_M T = 0 \quad\Leftrightarrow\quad
	\left(\cD^A \lambda^B -\Omega^{ABC}\lambda_C\right) \cD_A T = 0\ ,
\ee
must be satisfied.
Introducing Lorentz generators for $O(D,D)$ spinors $\Sigma^{AB} = \G^{AB}$, the Dirac operator reads
\be
	\G^A\nabla_A =
	\G^A\left( \cD_A -\frac{1}{2}\omega_{ABC}\G^{BC} \right) =
	\G^A\,\cD_A -\frac{1}{2}\G^A\,\omega^B{}_{BA}	- \frac{1}{2}\G^{ABC}\,\omega_{[ABC]}\ ,
\ee
such that it only involves the antisymmetric and trace parts of the connection, i.e. those determined by (\ref{SixthCondition}) and (\ref{SeventhCondition}) in terms of the fluxes.

~

 For our present purposes it is sufficient to consider the associated Dirac operator $\dop = \G^A\nabla_A$, for which only the generalized torsion condition and self-adjoint property matter. When these conditions hold, this operator reads
\be\label{rr_dop}
\begin{split}
    \dop &= \slashed \cD - \frac{1}{2}\slashed \cF_1 - \slashed \cF_3 \\
         &= \G^A \cD_A
         	-\frac{1}{2} \G^A \cF_A
             -\frac{1}{6}\G^{ABC} \cF_{ABC}\ .
\end{split}
\ee
A simple computation shows that this operator precisely reproduces $d+H\wedge$ on components when (\ref{beinparam}) is assumed. More generally, using the BI (\ref{Constraints}), this operator is nilpotent up to terms that vanish when the strong constraint holds
\be\label{rr_nil}
    \dop^2 = \partial^M\partial_M -\frac{1}{2}\Omega^A{}_{BC}\G^{BC}\cD_A -\cD^A d\,\cD_A
 	-\frac{1}{4}\cZ -\frac{1}{4}\cZ_{AB}\G^{AB} -\frac{1}{6}\cZ_{ABCD}\G^{ABCD}.
\ee
It is interesting to notice the appearance of $\cal Z$ here. We mentioned before that a constraint involving this combination of fluxes would arise in the maximal supergravity completion of the theory, so it was to be expected that it would arise in a type II formulation of the theory.
With a nilpotent operator that generalizes $d+H\wedge$ in our hands, we can easily rewrite (\ref{df_rfs}) in terms of the spinor $\cG$
\be\label{rr_fsn}
\begin{split}
    \cG &= \dop \cC\ ,\\
    \dop \cG &= 0\ ,
\end{split}
\ee
where the spinor $\cC$ plays the role of gauge potential. The field strength is then invariant under gauge transformations
\be\label{rr_gaugesymm}
    \d_\chi \cC = \dop \chi\ ,
\ee
parameterized by $\chi$, provided the strong constraint dependent condition
\be
	\dop^2 \chi = 0\ ,
\ee
is satisfied, where $\dop^2$ is given by (\ref{rr_nil}). Let us note that in a SS type compactifications, with $\cZ_{ABCD} = \cZ_{AB} = 0$ and with $\chi$ depending on external coordinates only, this condition further restrains the quantity $\cZ$ to be vanishing, in accordance with the known constraints for the embedding of $\cN=4$ in $\cN=8$ supergravity \cite{Aldazabal:2011yz}. The variation of the field strength under NS-NS generalized diffeomorphisms reads
\be
	\d_\xi \cG = \xi^M \, \partial_M \cG
				+ \G^N \partial^M \xi_N \, \partial_M \cG
				-\frac{1}{2} \slashed \Delta^{ns}_1
				- \slashed \Delta^{ns}_3\ , \label{dG}
\ee
where $\Delta^{ns}_{1,3}$ are the deviations from scalar behavior for $\cF_A$ and $\cF_{ABC}$ as read in (\ref{f3var}) and (\ref{f1var}). Therefore, for the field strength to transform as a scalar, the vanishing of the last three terms in (\ref{dG}) must be imposed as a constraint.

A pseudo-action for the R-R sector can then compactly be written as
\be\label{rr_action}
	S = -\frac{1}{4}\int d X e^{-2d}\, \overline\cG \Psi_+\cG\, ,
\ee
where $\overline\cG = \cG^T\,C$ and where $C$ is the charge conjugation matrix. Writing $\cG = \dop \cC$ and varying the potential $\cC$ in this action yields the equations of motion
\be\label{rr_eom}
	\dop \Psi_+ \cG=0\ ,
\ee
which are equivalent to the BI when the self-duality holds. Varying the bein in this action, with $\cG = \dop \cC$, and using the self-duality condition, one obtains the following modification to the bein equations of motion when RR fields are present
\be\label{rr_gemt}
	\cG^{RR}{}_{[AB]} = -\frac{1}{4}\overline{\cG}\G_{AB}\cG\ .
\ee
This pseudo-action does not contribute to the dilaton equation of motion.

It would be interesting to see if a SS compactification of the R-R sector reproduces the RR gaugings of gauged supergravity.

\subsection{Heterotic}

The inclusion of $n$ heterotic vectors $A_i{}^\alpha$ with $\alpha = 1,\dots,n$  in a duality covariant way
was done in \cite{heterotic} after \cite{Maharana:1992my},\cite{Siegel:1993th} (see also \cite{Andriot:2011iw}). One possibility is  to enlarge the global symmetry group to $G = O(D,D+n)$ with metric
 \be
 \eta_{MN} = \left(\begin{matrix} 0 & \delta^i{}_j & 0 \\ \delta_i{}^j & 0 & 0\\ 0& 0 & \delta_{\alpha \beta}\end{matrix}\right)\ , \ \ \ \ M,N = 1,\dots,2D+n\ .
 \ee
The bein can then be extended to include the vector fields as
\be
 {\cal E}^A{}_M  = \left(\begin{matrix}e_a{}^i &  e_{a}{}^k \left(B_{ki} - \frac{1}{2} A_k{}^\gamma A_{\gamma i}\right) & e_a{}^k A_{k\beta}\\ 0 & e^a{}_i & 0\\ 0 & A^\alpha{}_{i}& \delta^{\alpha}{}_{\beta}\end{matrix}\right)\, ,
\ee
and then all the covariant expressions in this paper just apply for these generalized quantities. We just mention this for completeness to highlight the fact that including vectors in this setup is straightforward, and for simplicity in this paper we will not  include vectors in the analysis. Interested readers can see how the vectors give rise to Maxwell fluxes
in \cite{Aldazabal:2011nj}
and to their corresponding BI in \cite{FreedWitten}.

%%%%%%%%%%%%%%%%%%%%%%%%%%%%%%%%%%%%%%%%%%%%%%%%%%%%%%%%%%%%%%%%%%%%%%%%%%%%%%%%%%%%%%%%%%%%%%%

\section{Bianchi identities}\label{BianchiId}

In the previous sections we have identified three quantities (\ref{gf_ktens2}), (\ref{gf_ktens4}) and (\ref{gf_ktens0}) that vanish under the strong constraint (\ref{strong}):
\bea
{\cal Z}_{ABCD} &=& -\frac{3}{4}\Omega_{E[AB}\Omega^E{}_{CD]}\, ,\label{Z4}\\
{\cal Z}_{AB} &=& \left(\partial^M\partial_M {\cal E}_{[A}{}^N \right){\cal E}_{B]N}
                    - 2\Omega^C{}_{AB} \cD_C d \, ,\label{Z2}\\
{\cal Z}&=&-2\cD^Ad\,\cD_A d+2\partial^M\partial_M d+\frac{1}{4}\Omega^{ABC}\Omega_{ABC} \, .\label{Z0}
\eea
They appeared when analyzing the symmetries, constraints and equations of motion. Interestingly, these quantities can be written purely in terms of fluxes and their derivatives. They lead to the following duality orbits of {\it generalized BI} for all the dual fluxes
\bea
   \cD_{[A}{\cal F}_{BCD]}-\frac{3}{4}{\cal F}_{[AB}{}^E {\cal F}_{CD]E}  &=&  {\cal Z}_{ABCD}  \, ,\label{FirstBI2}\\
    \cD^C {\cal F}_{CAB} +2\cD_{[A}{\cal F}_{B]}-{\cal F}^C {\cal F}_{CAB} &=&  {\cal Z}_{AB} \, ,\label{SecondBI2}\\
    \cD^A{\cal F}_A-\frac{1}{2}{\cal F}^A{\cal F}_A +\frac{1}{12}{\cal F}^{ABC}{\cal F}_{ABC} &=& {\cal Z} \, .\label{ThirdBI2}
\eea
When R-R fields are present, we find the additional  identity
\be
	\dop \cG = \cZ_{RR}\ ,
\ee
with
\be
	\cZ_{RR} =
	\left(\partial^M\partial_M -\frac{1}{2}\Omega^A{}_{BC}\G^{BC}\cD_A -\cD^A d\,\cD_A
 	-\frac{1}{4}\cZ -\frac{1}{4}\cZ_{AB}\G^{AB} -\frac{1}{6}\cZ_{ABCD}\G^{ABCD}\right)\cC\ .
\ee

%%%%%%%%%%%%%%%%%%%%%%%%%%%%%%%%%%%%%%%%%%%%%%%%%%%%%%%%%%%%%%%%%%%%%%%%%%%%%%%%%%%%%%%%%%%%%%%
\subsection{Relation to standard fluxes}

The fluxes ${\cal F}_{ABC}$ encode the standard T-dual fluxes. This can be seen by splitting the indices as
\be
{\cal F}_{abc} = H_{abc}\ , \ \ \ {\cal F}^a{}_{bc} = \tau_{bc}{}^a\ , \ \ \ {\cal F}^{ab}{}_c = Q_c{}^{ab}\ , \ \ \ {\cal F}^{abc} = R^{abc}\, .
\ee
Notice that being defined with planar indices these fluxes are T-duality invariant, but after a rotation with the generalized bein, they obey the usual T-duality chain
 \begin{equation}
H_{ijk}\ \ {\stackrel{T_k} {\longleftrightarrow}}\ \ \tau_{ij}{}^k\
\ {\stackrel{T_j} {\longleftrightarrow}}\ \ Q_i{}^{jk}\ \
{\stackrel{T_i} {\longleftrightarrow}}\ \ R^{ijk}\label{chainruleplanar}
\end{equation}
where T-dualities are defined by
\begin{equation}
(T_l)^{ N}{}_M = \delta^{  N}{}_M - \delta^{N,
l}\delta_{M, l} -  \delta^{N, l+D}\delta_{M,
l+D} +  \delta^{N, l+D}\delta_{M, l} +
\delta^{N,l}\delta_{M, l+D} \, .\label{TDualityTransf}
\end{equation}

Splitting in components equation (\ref{Z4}) we find

\bea
\cD_{[a}H_{bcd]} -\frac{3}{2}H_{e[ab} \tau_{cd]}{}^e &=&{\cal Z}_{abcd}\ ,\nn\\
3\cD_{[a}\tau_{bc]}{}^d - \cD^d H_{abc} +3\tau_{[ab}{}^e\tau_{c]e}{}^d
-3Q_{[a}{}^{de}H_{bc]e} &=& {\cal Z}_{abc}{}^d\ ,\nn\\
2\cD_{[a}Q_{b]}{}^{cd} +2 \cD^{[c}\tau_{ab}{}^{d]}
- \tau_{ab}{}^e Q_e{}^{cd} - H_{abe}R^{ecd} + 4Q_{[a}{}^{e[c}\tau_{b]e}{}^{d]} &=&{\cal Z}_{ab}{}^{cd}\ ,\label{obi}\\
3\cD^{[a}Q_d{}^{bc]} - \cD_d R^{abc} +3Q_e{}^{[ab}Q_d{}^{c]e}
-3\tau_{de}{}^{[a}R^{bc]e} &=& {\cal Z}^{abc}{}_d\ ,\nn\\
\cD^{[a}R^{bcd]} -\frac{3}{2}R^{e[ab} Q_e{}^{cd]} &=&{\cal Z}^{abcd}\ .\nn
\eea
From equation (\ref{Z2}) we get
\bea
\cD^c H_{abc} +\cD_c\tau_{ab}{}^c +2\cD_{[a}\cF_{b]}
-\cF^cH_{abc} - \cF_c\tau_{ab}{}^c&=&{\cal Z}_{ab}\ ,\nn\\
\cD^c \tau_{ca}{}^b +\cD_c Q_a{}^{bc} +\cD_{a}\cF^{b} - \cD^{b}\cF_{a}
-\cF^c \tau_{ca}{}^b - \cF_c Q_a{}^{bc}&=&{\cal Z}_a{}^b\ ,\\
\cD_c R^{abc} +\cD^c Q_c{}^{ab} +2\cD^{[a}\cF^{b]}
-\cF_c R^{abc} - \cF^c Q_c{}^{ab}&=&{\cal Z}^{ab}\ ,\nn
\eea
and equation (\ref{Z0}) reads in components
\be
\cD^a\cF_a + \cD_a\cF^a - \cF^a\cF_a +\frac{1}{6}H_{abc}R^{abc}
+ \frac{1}{2} \tau_{ab}{}^c Q_c{}^{ab} = {\cal Z}\ .
\ee

We can now use the following extended parameterization
\be
{\cal E}^A{}_M = \left(\begin{matrix} e_a{}^k & e_a{}^j B_{jk} \\ e^a{}_j
\beta^{jk} & e^a{}_k + e^a{}_i \beta^{ij}B_{jk}\end{matrix}\right)\, ,\label{VielbParam}
\ee
where a bi-vector $\beta^{ij}$ was introduced to get the most general bein.
With this parameterization  the fluxes match those computed in
\cite{Aldazabal:2011nj}, namely
\be\label{nf}
\begin{split}
{\cal F}_{abc} &= 3\left[ \nabla_{[a} B_{bc]}  - B_{d[a} \tilde\nabla^d B_{bc]} \right]\, ,\\
{\cal F}_{ab}{}^{c} &= 2 \Gamma_{[ab]}{}^c + \tilde \nabla^c B_{ab} + 2 \Gamma^{mc}{}_{[a} B_{b]m} + \beta^{cm}{\cal F}_{mab}\, ,\\
{\cal F}_c{}^{ab} &= 2 \Gamma^{[ab]}{}_c +\partial_c\beta^{ab}+  B_{cm} \tilde \partial^m\beta^{ab} + 2 {\cal F}_{mc}{}^{[a}\beta^{b]m}  -{\cal F}_{mnc} \beta^{ma}\beta^{nb}\, ,\\
{\cal F}^{abc} &= 3\left[\beta^{[\underline am}\nabla_m \beta^{\underline b\underline c]}
+ \tilde \nabla^{[a} \beta^{bc]} +
 B_{mn} \tilde \nabla^n\beta^{[ab} \beta^{c]m} + \beta^{[\underline am}\beta^{\underline bn} \tilde\nabla^{\underline
c]}B_{mn}
 \right ] +
\beta^{am}\beta^{ bn}\beta^{cl}{\cal F}_{mnl}\, ,
\end{split}
\ee
and
\bea
\mathcal{F}_{a}&=&-\tilde{\nabla}^c B_{ac} + \Gamma^{cd}{}_a B_{dc} - \Gamma_{ca}{}^{c} + 2 B_{ac}\tilde{\nabla}^c d + 2\nabla_{a}d\, ,\nn\\
\mathcal{F}^{a}  &=&-\Gamma^{ca}{}_{c}-\tilde{\nabla}^d\beta^{ac}B_{cd}-\Gamma^{da}{}_e\beta^{ec}B_{cd}-\beta^{ac}\tilde{\nabla}^dB_{cd}+ 2\tilde{\nabla}^a d +2\beta^{ac}B_{ce}\tilde{\nabla}^e d\nn\\
&&  +2\beta^{ac}\nabla_c d -\nabla_c\beta^{ac} + \Gamma_{cd}{}^a\beta^{dc} \, ,\nn
\eea
where we have used the following relations and definitions
\be
e_a{}^i e^a{}_j = \delta^i_j \ , \ \ \ \ \ \ e_a{}^i e^b{}_i = \delta_a^b\, ,\quad
B_{ab} = e_a{}^i e_b{}^j B_{ij} \ , \ \ \ \ \ \ \beta^{ab} = e^a{}_i e^b{}_j  \beta^{ij}\, ,\nn
\ee
\be
\partial_a = e_a{}^i \partial_i  \ , \ \ \ \ \ \ \tilde \partial^a = e^a{}_i \tilde \partial^i\, ,\nn
\ee
\bea
\nabla_a B_{bc}=\partial_a B_{bc}-\Gamma_{ab}{}^dB_{dc}-\Gamma_{ac}{}^dB_{bd}\, ,\quad
\tilde \nabla^a B_{bc}=\tilde\partial^aB_{bc}+\Gamma^{ad}{}_{b}B_{dc}+\Gamma^{ad}{}_{c}B_{bd}\, ,\nn\\
\nabla_a \beta^{bc}=\partial_a\beta^{bc}+\Gamma_{ad}{}^b\beta^{dc}+\Gamma_{ad}{}^c\beta^{bd}\, ,\quad
\tilde \nabla^a \beta^{bc}=\tilde\partial^a\beta^{bc}-\Gamma^{ab}{}_{d}\beta^{dc}-\Gamma^{ac}{}_{d}\beta^{bd}\, ,\nn
\eea
and
\be
\Gamma_{ab}{}^c = e_a{}^i \partial_i e_b{}^j e^c{}_j \ , \ \ \ \ \ \  \Gamma^{ab}{}_c = e^a{}_i \tilde \partial^i e^b{}_j e_c{}^j\, .\label{Gammas}
\ee

After imposing the strong constraint and selecting the frame  $\tilde \partial^i = 0$, the fluxes (\ref{nf})
agree with those obtained in \cite{Halmagyi:2009te, Blumenhagen:2012pc}, namely
\be\label{fb}
\begin{split}
{\cal H}_{abc} &= 3\left[ \partial_{[a} B_{bc]} + f_{[ab}{}^d B_{c]d} \right] \equiv 3\nabla_{[a} B_{bc]} \, ,\\
{\cal F}_{ab}{}^{c}&=f_{ab}{}^c-{\cal H}_{abm}\beta^{mc}\, ,\\
 {\cal Q}_c{}^{ab}& = \partial_c\beta^{ab} + 2f_{cm}{}^{[a} \beta^{mb]} +{\cal H}_{cmn} \beta^{ma}\beta^{nb} \, , \\
{\cal R}^{abc} &= 3\left[\beta^{[am}\partial_m \beta^{bc]} + f_{mn}{}^{[a} \beta^{bm}\beta^{c]n}\right] -
{\cal H}_{mnp} \beta^{ma}\beta^{nb}\beta^{pc} \, ,
\end{split}
\ee
where $f_{ab}{}^c = 2\Gamma_{[ab]}{}^c$. Applying the same restrictions on (\ref{obi}),
 the resulting equations  exactly match
 the BI  derived in \cite{Blumenhagen:2012pc}
(recall that the right hand sides of (\ref{obi})
vanish when
the strong constraint is imposed).

%%%%%%%%%%%%%%%%%%%%%%%%%%%%%%%%%%%%%%%%%%%%%%%%%%%%%%%%%%%%%%%%%%%%%%%%%%%%%%%%%%%%%%%%%%%%%%%

The fluxes (\ref{fb}) were shown
to be the coefficients of
the following Roytenberg algebra:
\be
\begin{split}
\left [e_a,e_b\right ]&={\cal F}_{ab}{}^ce_c+{\cal H}_{abc}e^c\, ,\\
\left [e_a,e^b\right ]&={\cal Q}_{a}{}^{bc}e_c-{\cal F}_{ac}{}^be^c\, ,\\
\left [e^a,e^b\right ]&={\cal Q}_{c}{}^{ab}e^c+{\cal R}^{abc}e_c\, ,
\end{split}\label{roy}
\ee
obtained as
a Courant algebroid on basis sections
$\{e_a,e^b\}\in TM\oplus T^*M$ in   \cite{Halmagyi:2009te, gmpw, Blumenhagen:2012pc}.
And they also determine the   Jacobiators
\bea
{\rm Jac}(e_a,e_b,e_c)&=&\frac 12{\cal D}{\cal H}_{abc}\, ,\nn\\
{\rm Jac}(e_a,e_b,e^c)&=&\frac 12{\cal D}{\cal F}_{ab}{}^c\, ,\nn\\
{\rm Jac}(e_a,e^b,e^c)&=&\frac 12{\cal D}{\cal Q}_{a}{}^{bc}\, ,\nn\\
{\rm Jac}(e^a,e^b,e^c)&=&\frac 12{\cal D}{\cal R}^{abc}\, ,\label{royj}
\eea
with ${\cal D}=d^H+d_\beta^H$, $d^H$ and $d_\beta^H$  being the $H$-twisted de Rham and
 Poisson differentials respectively, which hold up to the BI
(see \cite{Blumenhagen:2012pc} for details).

Here we notice that DFT provides a natural framework containing these structures covariantly.
Indeed, a covariant expression encoding the algebra (\ref{roy})
follows from the
C-bracket of generalized beins:
\be
\left [{\cal E}_A{}^M,{\cal E}_B{}^N\right  ]^{(C)}_P=
{\cal F}_{ABC}{\cal E}^{C}{}_P\, ,\label{croy}
\ee
and the cyclic
sum of double C-brackets  gives:
\be
\left [[{\cal E}_A{}^M,{\cal E}_B{}^N]^{(C)},{\cal E}_C{}^P\right ]_Q^{(C)} + {\rm cyclic} = -4
{\cal Z}_{ABCE}{\cal E}^E{}_Q+\frac 12{\cal D}_E{\cal F}_{ABC}{\cal E}^E{}_Q\, ,\label{cbi}
\ee
precisely  the covariant generalization of (\ref{royj}).

\subsection{Towards a first order formulation of DFT}

In the usual description of supergravity, magnetic sources appear as defects in the BI of the field strengths of the theory. For instance, for an NS5-brane one has
\be
	dH = T_{NS5}\, \d_4\ ,
\ee
where $\d_4$ is a delta function four-form based on the brane's worldvolume, with legs in the directions transverse to the worldvolume. In this picture the three-form cannot be defined globally from the two-form gauge field. Adding a Lagrange multiplier six-form, the sourceless BI follows as an equation of motion from
\be
	S = \int \left(-\frac{1}{2} \star H\wedge H - B_6\wedge dH\right)\ ,
\ee
where the three-form is now treated as independent of $B_2$ and one has two first-order equations of motion.
Adding to this action a Wess-Zumino coupling on the NS5-brane worldvolume
\be
	S_{WZ} = T_{NS5}\int_{{\cal W}_6} \pi_{{\cal W}_6} (B_6) = T_{NS5}\int \d_4\wedge B_6\ ,
\ee
one precisely recovers the BI for the three-form in presence of an NS5-brane, as the equation of motion of $B_6$. One can then integrate $H$ out and express the dynamics in terms of $B_6$ solely\footnote{This is due to the linear nature of this action. When non-linearities are present,  for instance like the Chern-Simmons term of eleven-dimensional supergravity, one can in general not get rid of the electric potential.}.

Since $dH=0$ is contained in our BI and since $dH \neq 0$ when an NS5-brane is present, the generalized BI cannot hold as such when sources are present. This in turn suggests that the generalized diffeomorphisms themselves should be corrected, but this lies beyond the scope of this paper. We propose that a flux configuration in the presence of some extended objects satisfies
\bea
    \cD_{[A}{\cal F}_{BCD]}-\frac{3}{4}{\cal F}_{[AB}{}^E {\cal F}_{CD]E} &=& \cJ_{ABCD}\ , \label{BI4}\\
    \cD^C {\cal F}_{CAB} +2\cD_{[A}{\cal F}_{B]}-{\cal F}^C {\cal F}_{CAB} &=& \cJ_{AB}\ , \label{BI2}\\
    \cD^A{\cal F}_A-\frac{1}{2}{\cal F}^A{\cal F}_A +\frac{1}{12}{\cal F}^{ABC}{\cal F}_{ABC}
    &=& \cJ\ ,\label{BI0}\\
    \dop\cG &=& \cJ_{RR} \ ,\label{BIRR}
\eea
where $\cJ_{\dots}$ represent currents for these (postulated) extended objects and where, for instance, $\cJ_{RR}$ represents a D-brane current. For simplicity, we assume through this section that the strong constraint terms $\cZ_{\dots}$ are vanishing. We however want to stress that, since the quantities $\cZ_{\dots}$ enter the BI on the same footing as the currents $\cJ_{\dots}$, it seems that one has {\it a-priori} the option to describe an extended object either by a source term $\cJ_{\dots} \neq 0$ or by a strong constraint-violating solution with $\cZ_{\dots} \neq 0$. For non-vanishing currents, the fluxes cannot be given any longer in terms of the bein and dilaton. We can however introduce deviation terms and write them as
\bea
	\cF_{ABC} &=& f_{ABC}(\cE) + \Theta_{ABC}\ ,\\
	\cF_{A} &=& f_{A}(\cE, d) + \Theta_{A}\ ,\\
	\cG	 &=& \dop\cC + \Theta_{RR}\ ,
\eea
where $f_{ABC} = 3\Omega_{[ABC]}$ and $f_A = 2\cD_A + \Omega^B{}_{BA}$. Plugging these general expressions in the sourced BI yields
\bea
	\nabla^f_{[A}\Theta_{BCD]} -\frac{3}{4}\Theta_{[AB}{}^E\Theta_{CD]E} &=& \cJ_{ABCD}\ ,\\
	2\nabla^f_{[A}\Theta_{B]}+ \left( \cD^C - f^C\right)\Theta_{CAB}
	+\Theta^C \Theta_{CAB} &=& \cJ_{AB}\ , \\
	\left( \cD^A - f^A\right)\Theta_{A} - \frac 12 \Theta^A\Theta_A +\frac 1{12}\left (2f^{ABC}+\Theta^{ABC}\right )\Theta_{ABC} &=&{\cal J}\ ,
\eea
where the connection in the pseudo-covariant derivative
\be
\nabla^f_A\Theta_B = {\cal D}_A\Theta_B- \omega_{AB}{}^C\Theta_C\ ,
\ee
satisfies the following conditions
\bea
	\omega_{[AB]C} &=& \frac 12 f_{ABC}\ , \\
	\omega^B{}_{BA} &=& f_A\ .
\eea
Let us note that the vanishing of the currents does not imply in principle the vanishing of the deviation terms, but instead yields complex non-linear differential equations.

We would now like to see if a first-order formulation of DFT is available in order to formulate couplings to magnetic objects from a dynamical perspective. A first-order formulation of the theory was first presented in \cite{Siegel:1993th}, with the spin connection treated as an independent variable determined by its equation of motion. Following the previous reasoning employed for coupling the NS5-brane to the three-form, we introduce an antisymmetric Lagrange multiplier 4-tensor $B^{ABCD}$ imposing the first BI as its equation of motion, and consider the fluxes as independent variables. The modified action reads
\be\begin{split}
	S' = \int dX e^{-2d}\ \Bigg[&2\cD^{\check{A}}\cF_A - \cF^{\check{A}}\cF_A
		+\frac{1}{6} \breve{\cF}^{ABC} \cF_{ABC} -2 \cJ \\
	+ &B^{ABCD}\left(\cD_{A}{\cal F}_{BCD}
	-\frac{3}{4}{\cal F}_{AB}{}^E {\cal F}_{CDE} - \cJ_{ABCD} \right) \Bigg]
	+ S_{loc}\left(\cE, d\right),
\end{split}\ee
where we used the check notation (\ref{Check}) to indicate that indices are contracted with the planar generalized metric, and we defined (see (\ref{cupdefinitions}))
\be
\breve{\cF}^{ABC}  = \breve{S}^{ABCDEF} {\cF}_{DEF}\ .
\ee
The fluxes $\cF_{ABC}$ and $\cF_A$ are now treated as independent variables, the bein then enters the action only through derivatives $\cD_A$ and possibly the additional local action $S_{loc}$. Note also that (\ref{BI0}) has been used to rewrite the flux terms that vanish in the standard case when the strong constraint holds. Varying with respect to the various fields yields
\bea
	\d \cF_A &:\quad & \cF_A = f_A\ , \label{foe1} \\
	\d \cF_{ABC} &:\quad & \breve{\cF}^{ABC} =  3\left (\left({\cal D}_D - f_D\right)B^{DABC}-\frac 32{\cal F}_{DE}{}^A
B^{DEBC}\right)\ , \label{foe2} \\
	\d B^{ABCD} &:\quad &
	 \cD_{[A}{\cal F}_{BCD]}-\frac{3}{4}{\cal F}_{[AB}{}^E {\cal F}_{CD]E} = \cJ_{ABCD}\ ,\label{foe3}\\
	\d \cE_A{}^M &:\quad & 2\cD^{[A}\cF_C S^{B]C} + B^{CDE[A}\cD^{B]}\cF_{CDE}
	= \cG_{loc}^{[AB]}\ , \label{foe4} \\
	\d d &:\quad &  2\cD^{\check{A}}\cF_A - \cF^{\check{A}}\cF_A
		+\frac{1}{6} \breve{\cF}^{ABC} \cF_{ABC} = 2 \cJ - S_{loc}
		+ \frac 12 \frac{\d S_{loc}}{\d d}\ , \label{foe5}
\eea
where the BI (\ref{foe3}) has already been used to simplify the dilaton equation of motion (\ref{foe5}). The equation of motion for $\cF_A$ (\ref{foe1}) automatically sets it to the standard value $f_A = 2\cD_A d - \Omega^B{}_{BA}$. Let us note that it is not clear that this action gives the correct equations of motion for dynamical fluxes in the presence of sources, but must only be considered as a first step toward such a description. Imposing by hand the relation $\cF_{ABC} = f_{ABC}$, the source $\cJ_{ABCD}$ has to vanish due to (\ref{foe3}) and (\ref{foe2}) can be rewritten as
\be
	\breve{\cF}^{ABC} = 3\nabla^f_D B^{DABC}\ .
\ee
Taking another divergence of this equation, we obtain
\be
	\nabla^f_C \breve{\cF}^{CAB} = - 3\nabla^f_C\nabla^f_D B^{CDAB}
	 = B^{CDE[A}\cD^{B]} \cF_{CDE}\ ,
\ee
where we dropped strong constraint-violating terms in the last equality. Combining with (\ref{foe4}), one then recovers the standard equations for DFT
\be
	2\cD^{[A}\cF_C S^{B]C} + \nabla^f_C \breve{\cF}^{CAB} = \cG_{loc}^{[AB]}\ ,
\ee
up to the local source term $\cG_{loc}^{[AB]}$ and up to strong constraint-vanishing terms. Using (\ref{foe1}) and the assumption $\cF_{ABC} = f_{ABC}$, the dilaton equation of motion is then also recovered from (\ref{foe5})
\be
	\cR = 2 \cJ - S_{loc}
		+ \frac 12 \frac{\d S_{loc}}{\d d}\ ,
\ee
again up to source and strong constraint-vanishing terms. It would be interesting to pursue this study with, for instance, other Lagrange multipliers to take into account all possible sources.

%%%%%%%%%%%%%%%%%%%%%%%%%%%%%%%%%%%%%%%%%%%%%%%%%%%%%%%%%%%%%%%%%%%%%%%%%%%%%%%%%%%%%%%%%%%%%%%%%%%%%%%%%%%%%%%%%%%%%%%%

%%%%%%%%%%%%%%%%%%%%%%%%%%%%%%%%%%%%%%%%%%%%%%%%%%%%%%%%%%%%%%%%%%%%%%%%%%%%%%%%%%%%%%%%%%%%%%%
%%%%%%%%%%%%%%%%%%%%%%%%%%%%%%%%%%%%%%%%%%%%%%%%%%%%%%%%%%%%%%%%%%%%%%%%%%%%%%%%%%%%%%%%%%%%%%%
%%%%%%%%%%%%%%%%%%%%%%%%%%%%%%%%%%%%%%%%%%%%%%%%%%%%%%%%%%%%%%%%%%%%%%%%%%%%%%%%%%%%%%%%%%%%%%%

%%%%%%%%%%%%%%%%%%%%%%%%%%%%%%%%%%%%%%%%%%%%%%%%%%%%%%%%%%%%%%%%%%%%%%
\subsection{Including sources}

Since T-duality exchanges Dirichlet and Neuman boundary conditions in the open string sector, it
connects D-branes of different dimensionalities, and the full T-duality orbits of D-branes have been nicely encoded in the double space in \cite{Doublebranes}.
Here instead, we will focus on NS-NS branes lying in the orbit of the NS5-brane and KK5-monopole, along the lines of \cite{deBoer:2010ud} and \cite{Hassler:2013wsa}. It is known that these two configurations are related by  T-duality, and that they are not sufficient to span the full duality orbit.

The study of exotic brane orbits is closely related to that of non-geometric fluxes. To picture the idea, one can start with a two-form flux background $H_{ijk}$ and T-dualize it to a twisted torus, characterized by a geometric flux $\tau_{ij}{}^k$. Additional T-dualities lead to the more exotic non-geometric fluxes $Q_i{}^{jk}$ and $R^{ijk}$ through the chain (\ref{chainruleplanar}).
The backgrounds generating these fluxes have very different topologies, characterized by the T-duality elements needed to glue coordinate patches after undergoing monodromies. In the $H$-flux background, the patches are connected through gauge transformations of the two-form, and in the $\tau$-background the transition functions are diffeomorphisms. More generally, the $Q$-background makes use of the T-duality group, and is therefore called a T-fold \cite{doublegeom}.

The NS5-brane carries a non-constant $H$-flux and the KK5-monopole has a non-constant $\tau$-flux, so they correspond to $H$- and $\tau$-flux
backgrounds respectively. The next object in the T-duality chain, the $Q$-brane \cite{Hassler:2013wsa}, will carry a non-constant $Q$-flux and will therefore be a T-fold.
The $R$-brane would be the last  object in the chain. The aim of this subsection is to study some properties of these dual objects.  In the presence of sources the BI locally breakdown on the world-volume, so we will use the duality orbits of BI  to speculate about brane orbits.

Before we begin, let us introduce two ``frames'' in which geometric and non-geometric backgrounds are best described. For a recent detailed analysis we refer to \cite{Blumenhagen:2013aia}.
%%%%%%%%%%%%%%%%%%%%%%%%%%%%%%%%%%%%%%%%%%%%%%%%%%%%%%%%%%%%%%%%%%%%%%%%%%%%%%%%%%%%%%%%%%%%%%%
\subsubsection*{Geometric versus non-geometric frames}
We have been completely general in parameterizing the generalized bein as an $O(D,D)$ element
\be
{\cal E}^A{}_M = \left(\begin{matrix} e_a{}^k & e_a{}^j B_{jk} \\ e^a{}_j
\beta^{jk} & e^a{}_k + e^a{}_i \beta^{ij}B_{jk}\end{matrix}\right)\, ,\label{VielbParam}
\ee
in terms of a $D$-dimensional bein $e_a{^i}$, a two-form $B_{ij}$ and an antisymmetric bi-vector $\beta^{ij}$. For this parameterization the generalized metric takes the form
\be
{\cal H}_{MN} = \left(\begin{matrix} g^{ij} -\beta^{im} g_{mn}\beta^{nj} &  & (g^{ik} -\beta^{im} g_{mn}\beta^{nk})B_{kj} - \beta^{im}g_{mj}\\ & & \\B_{ik}(\beta^{km} g_{mn}\beta^{nj}-g^{kj} ) + g_{im} \beta^{mj} & & {\begin{matrix} g_{ij} - B_{ik}(g^{kl} - \beta^{km}g_{mn} \beta^{nl})B_{lj} \\+ \ g_{im}\beta^{mn}B_{nj}  + B_{im}\beta^{mn}g_{nj} \end{matrix}}\end{matrix}\right)\, .\label{GenMetricParam}
\ee

Given that the generalized bein belongs to the coset $G/H$, defined in this way it is over-parameterized. Only $D^2$ degrees of freedom are physical, while the remaining $D(D-1)$ can be removed through a gauge choice. For example, for the {\it geometric } configurations defined in terms of a $B$-field and a metric, it is better to remove the $\beta$-dependence through a $O(1,D-1)^2$ transformation. On the other hand, there are {\it non-geometric} configurations for which it is better to remove the $B$-field, and describe the background in terms of $\beta$. We will therefore refer in what follows to two different gauge choices or {\it frames}. Also, given that the configurations we will consider will be locally geometric, the strong constraint will be automatically satisfied in this section, and we will choose the $\tilde \partial^i = 0$ T-duality frame, in which the fluxes reduce to (\ref{fb}).

\subsubsection*{Geometric frame}
The geometric frame corresponds to the gauge choice $\beta^{ij} = 0$ and the generalized metric reads
\be
{\cal H}_{MN} = \left(\begin{matrix} g^{ij} &   g^{ik} B_{kj}  \\-B_{ik} g^{kj}  &  g_{ij} - B_{ik} g^{kl} B_{lj}\end{matrix}\right)\, .\label{GenMetricGeom}
\ee
This is the frame usually considered for geometric descriptions of supergravity backgrounds described in terms of a $B$-field and a metric. The corresponding three-form $H_{ijk}$ and the geometric flux $\tau_{ij}{}^k$ in curved and planar indices  read
\bea
{\cal H}_{abc} &=& 3\left[ \partial_{[a} B_{bc]} + f_{[ab}{}^d B_{c]d} \right]  \, ,\nn\\
{\cal F}_{ab}{}^{c}&=&f_{ab}{}^c\, , \ \ \ \ \ \
 {\cal Q}_c{}^{ab} = 0 \ , \ \ \ \ \ {\cal R}^{abc} \ =\ 0\ ,
\eea
and
\bea
H_{ijk} &=& e^a{}_ie^b{}_je^c{}_k {\cal H}_{abc} = 3\partial_{[i}B_{jk]}\ ,\nn \\
\tau_{ij}{}^k &=& e^a{}_ie^b{}_j e_c{}^k {\cal F}_{ab}{}^c  = 2 \Gamma_{[ij]}{}^k\ , \ \ \ \ \ \Gamma_{ij}{}^k  = \partial_i e_a{}^k e^a{}_j\ ,\nn\\
Q_i{}^{jk} &=& 0 \ , \ \ \ \ \ R^{ijk} \ = \ 0 \ ,\label{curvedgeometric}
\eea
respectively. The dilaton flux can be written as
\be
f_i = e^a{}_i {\cal F}_a = 2 \partial_i \phi + \tau_{ij}{}^j\, .
\ee
The only non-trivial BI from the previous section then read (see Appendix \ref{GR})
\bea
\partial_{[i}H_{jkl]} &=& {\cal J}_{ijkl}\ ,\\
-3 R^l{}_{[ijk]} = \nabla_{[i} \tau_{jk]}{}^l + \tau_{[ij}{}^m \tau_{k]m}{}^l &=& {\cal J}_{ijk}{}^l\ ,\\
2 R_{[ij]} +4 \partial_{[i}\partial_{j]}\phi = \nabla_k \tau_{ij}{}^k + 2\partial_{[i}f_{j]} &=& {\cal J}_{ij}\ , \eea
where the ${\cal J}$ are only non-trivial on the world-volume of sources, as we will see later. Notice that ${\cal J}_{ij}$ sources a dilaton-like BI $df_i = 0$.
\subsubsection*{Non-geometric frame}
On the other hand, one can also define a non-geometric frame taking $B_{ij} = 0$ with generalized metric
\be
{\cal H}_{MN} = \left(\begin{matrix} g^{ij} -\beta^{im} g_{mn}\beta^{nj}   &  - \beta^{im}g_{mj} \\ g_{im} \beta^{mj} &  g_{ij}\end{matrix}\right)\, .\label{GenMetricNonGeom}
\ee
This frame was also considered in the context of DFT, and a differential geometry was considered for this frame in \cite{Andriot:2011uh}.
The fluxes in planar indices read
\bea
{\cal H}_{abc} &=& 0 \ , \ \ \ \ {\cal F}_{ab}{}^{c}\ = \ f_{ab}{}^c \ ,\nn\\
 {\cal Q}_c{}^{ab}& =& \partial_c\beta^{ab} + 2f_{cm}{}^{[a} \beta^{mb]}  \, , \nn\\
{\cal R}^{abc} &=& 3\left[\beta^{[am}\partial_m \beta^{bc]} + f_{mn}{}^{[a} \beta^{bm}\beta^{c]n}\right]  \ ,
\eea
while in curved indices they take the form
\bea
H_{ijk} &=& 0 \ , \ \ \ \ \tau_{ij}{}^k = 2 \Gamma_{[ij]}{}^k \ ,\nn\\
Q_i{}^{jk} &=& e^a{}_i e_b{}^je_c{}^k {\cal Q}_a{}^{bc}\ =\ \nabla_i \beta^{jk} + 2 \tau_{li}{}^{[j}\beta^{k]l} \ ,\nn \\
R^{ijk} &=& e_a{}^i e_b{}^je_c{}^k {\cal R}^{abc}\ =\ 3 \beta^{[\underline{i}l}\nabla_l \beta^{\underline{jk}]}\ .
\eea

\subsubsection{NS5-brane}

Let us briefly review here how the source term arises in the world-volume of an NS5-brane.  We begin by stating the solution in spherical coordinates \cite{Jensen:2011jna} on the transverse space, in the geometric frame
\be
ds^2 =  f(r) (dr^2 + r^2 d\theta^2 + r^2 \sin^2\theta  d\varphi^2 + d \psi^2) \ , \ \ \ \ \ H_{ijk} = \varepsilon_{ijk}{}^l \partial_l \ln f(r) \ ,\label{NS5}
\ee
where we are using the convention $\varepsilon_{r\theta\varphi\psi} = e$. We have omitted the world-volume coordinates since they play no role in the analysis. The brane is localized at $r = 0$, and the direction $\psi$ is just a circle over which the brane is smeared, so we take the warp factor as independent of this direction
\be
f(r) = 1 + \frac{m}{r}\, .
\ee

Since $H_3 = \ast_4 d \ln f(r)$, we have
\be
\star_4 d H_3 = \star_4 d\star_4 d \ln f(r) = \Delta \ln f(r) = \frac{1}{e} \partial_i \left(e g^{ij} \partial_j \ln f(r)\right) = 0 \ , \ \ \ \ \ \ {\rm at} \ \ r > 0\, .
\ee
However, when this quantity is integrated on a ball $V_a$ of arbitrary radius $r = a$ one obtains\footnote{We proceed as follows \bea \int d H_3 &=& 2\pi \int_{V_a} \star d H_3 dV = 2\pi \int_{ S_a}\frac{\partial f}{\partial r} dS  = 2\pi . (4\pi a^2) \left(-\frac{m}{a^2}\right) = - 8 \pi^2 m \eea where in the first step we integrated on $d\psi$ and in the second one we used Gauss' theorem. The result is independent of $a$, so $\ast_4 d H_3$ must be proportional to $\delta(r)$.}
\be
\int_{V_a} d H_3 = - 8 \pi^2 m\, .
\ee
Therefore, we are forced to conclude that
\be
\star_4 dH_3 = - 8 \pi^2 m \delta(r) = {\cal J}_{r\theta\varphi \psi}\, ,
\ee
and so the BI fails to hold on the world-volume of the brane. From the flux
\be
H_{\theta\varphi\psi} = - r^2 f \sin \theta \partial_r \ln f(r) \ , \ \ \ \ \ [H_{\theta\varphi\psi}]_{r>0} = m \sin\theta\, ,
\ee
we can define  a two-form field in the geometric frame
\be
B_{\varphi\psi} = m + \cos\theta r^2 f \partial_r \ln f \ , \ \ \ \ \ [B_{\varphi\psi}]_{r>0} = m (1 - \cos \theta)\, .
\ee

In order to make contact with a co-dimension two NS5 brane, we proceed as in \cite{deBoer:2010ud} compactifying in a base direction and then smearing it. Due to the compactification, it is better to implement a cylindrical coordinate system ($r, \theta, \varphi \to \rho, \vartheta, z$). The warp factor now takes the form
\be
f \to \sigma \log \frac{\mu} \rho \ , \ \ \ \ \sigma = \frac{\alpha'}{2\pi R_{z} R_\psi}\, ,
\ee
where  $\mu$ corresponds to a cut-off scale. Beyond this scale, the solutions fail to be trustable because co-dimension two objects cannot stand alone, but should rather form bound states through suitable superpositions. The parameter $\mu$ is then related to the distance between the NS5 brane and some other source, as explained in \cite{deBoer:2010ud}.

The solution now reads
\be
ds^2 = f(\rho)(d\rho^2 + \rho^2d\vartheta^2 + dz^2 + d\psi^2) \ , \ \ \ \ \ H_{ijk} = \varepsilon_{ijk}{}^l \partial_l \ln f(\rho)\ ,
\ee
and taking $\varepsilon_{\rho \vartheta z \psi} = e = f^2 \rho$ we find
\be
H_{\vartheta z \psi} = \sigma \ , \ \ \ \ B_{z\psi} = \sigma \vartheta \ \ \ \ \ \ {\rm at\ } \ \rho > 0\ .
\ee
In the new coordinate system we have
\be
{\cal J}_{\rho \vartheta z \psi} = - 8 \pi^3 \sigma \delta(\rho) \ .
\ee

Under a $\vartheta$-monodromy $\vartheta \to \vartheta + 2 \pi$, the $B$-field jumps $B_{z\psi} \to B_{z\psi} + 2\pi\sigma$, and plugging this in the generalized metric in the geometric frame we find
\be
{\cal H} (\vartheta + 2\pi) = \Omega_{NS}^T {\cal H}(\vartheta) \Omega_{NS}\ ,  \ \ \ \ \ \Omega_{NS}=\left(\begin{matrix} \mathbf{1}_4 & B(\vartheta = 2\pi) \\ 0 & \mathbf{1}_4 \end{matrix}\right)\, .
\ee
This is why $H$, the curvature for the $B$-field, receives a flux contribution. The matrix $\Omega_{NS}$ is an $O(2,2)$ element, and can be interpreted as a charge.

\subsubsection{KK5-monopole}

T-dualizing the previous solution in the direction $\psi$, we arrive at the co-dimension two $KK5$-monopole configuration, reading
\be
ds^2 = f (d\rho^2 + \rho^2 d\vartheta^2 + dz^2) + f^{-1}(d\psi - B^{(NS)}_{z\psi} dz)^2 \ , \ \ \ \ \sigma = \frac{R_\psi}{2\pi R_z}\, .\label{Cod2KKM}
\ee
As explained in \cite{Villadoro:2007tb}, this object now sources the metric BI: ${\cal J}_{\rho\vartheta z}{}^\psi$. In the context of compactifications, this sourcing translates into a relaxation of some quadratic constraints in half-maximal supergravities \cite{Villadoro:2007tb}, breaking ${\cal N} = 4 \to 2$. The charge of this object is also an element of $O(2,2)$, but now instead of corresponding to a $B$-transformation, it corresponds to a transformation of the form
\be
{\cal H} (\vartheta + 2\pi) = \Omega_{KK}^T {\cal H}(\vartheta) \Omega_{KK}\ ,  \ \ \ \ \ \Omega_{KK}=\left(\begin{matrix} e(\vartheta = 2\pi)^{-1} &0 \\ 0 & e(\vartheta = 2\pi)^{T} \end{matrix}\right)= T_\psi^{-1} \Omega_{NS} T_{\psi}\, .
\ee
Since now the vielbein $e$ jumps as $e_z{}^\psi \to e_z{}^\psi + 2\pi \sigma$ under a monodromy $\vartheta\to \vartheta + 2\pi$, the $\tau$ flux ( its ``curvature'') is turned on.

\subsubsection{$5^2_2$ brane}

Codimension-two branes have recently received renewed attention, in the context of exotic branes \cite{deBoer:2010ud}. There, starting with the KK5 solution (\ref{Cod2KKM}), a further T-duality is performed in the $z$-direction. The resulting object is a $Q$-background named $5^2_2$, which in the geometric frame  reads
\be
ds^2 = f (d\rho^2 + \rho^2 d\vartheta^2) + f K^{-1} (dz^2 + d\psi^2) \ , \ \ \ \ B_{z\psi} = - \sigma \vartheta K^{-1}\, , \ee
with \be K = f^2 + \sigma^2 \vartheta^2 \ , \ \ \ \ \sigma = \frac{R_\psi R_z}{2\pi \alpha'}\, .
\ee
However, as argued before, given that this is a non-geometric background, the  non-geometric frame seems more convenient to express this solution
\be
ds^2 = f (d\rho^2 + \rho^2 d\vartheta^2) + f^{-1} (dz^2 + d\psi^2)\ , \ \ \ \ \beta^{z\psi} = \sigma \vartheta\, .
\ee

Now, plugging this solution into the generalized metric (\ref{GenMetricNonGeom}), we see that under a monodromy $\vartheta \to \vartheta + 2\pi$, the fields mix through a $\beta$-transformation
\be
{\cal H} (\vartheta + 2\pi) = \Omega_{5^2_2}^T {\cal H}(\vartheta) \Omega_{5^2_2}\ ,  \ \ \ \ \ \Omega_{5_2^2} = \left(\begin{matrix} \mathbf{1}_4 & 0 \\ \beta(\vartheta = 2\pi) & \mathbf{1}_4 \end{matrix}\right) = T_z^{-1} \Omega_{KK} T_z\, ,
 \ee
where the $\beta$ field is shifted as $\beta^{z\psi} \to \beta^{z \psi} + 2\pi \sigma$. Therefore, the ``curvature'' of this field, namely the $Q$-flux, is non-vanishing as expected for a $Q$-brane
\be
Q_\vartheta{}^{z\psi} = \sigma\, .
\ee
It is then natural to assume that now it is the $dQ = 0$ BI ${\cal J}_{\rho \vartheta}{}^{z \psi}$ which is sourced on the worldvolume of the $5^2_2$ brane.

\subsection{Duality orbits of (exotic) branes }

Following the logic in \cite{Shelton:2005cf}, one could now proceed further, and T-dualize in some non-isometric direction. Now the solution will depend on a dual coordinate, and its geometric interpretation breaks down even locally, from a $D$-dimensional perspective. In DFT, this is not a problem, given that the notion of T-duality is generalized and allows for such kind of transformations. Given that the equations of motion are T-duality invariant, the
configuration  obtained in this way
will automatically solve them.

Such a configuration will however correspond to a particular representative of the orbit containing the branes that we explored in this section. In this sense, by construction, it can be T-dualized to a geometric object. Even more interesting is to determine if there exist truly non-geometric bound states of branes, belonging to truly non-geometric orbits. These cannot be T-dualized to a frame in which the configuration becomes geometric. A possibility is to consider bound states combining the presence of geometric and non-geometric branes, such that under T-dualities their roles get exchanged, but non-geometry is conserved. A first step in this direction was nicely achieved in \cite{Hassler:2013wsa}, were intersections of Q and R-branes were analyzed.

Non-geometric duality orbits were addressed for fluxes in \cite{Dibitetto:2012rk}. There, it was shown that genuine non-geometric orbits exist for fluxes, in which all types of gaugings $H$, $\tau$, $Q$ and $R$ are turned on simultaneously, and there is no T-duality frame in which any of them vanish. For such configurations the strong constraint must necessarily be relaxed, and it would be nice to explore whether this situation is reproduced by branes as well.

One can also consider the other duality orbits of BI and their associated sources ${\cal J}_{AB}$ and ${\cal J}$. To see what kind of objects they might be related to, it is instructive to analyze those for ${\cal J}_{ABCD}$. The two-form $B_2$ couples to the string $F_1$, and is dual to $B_6$ which is sourced by the NS5. Then, the NS5 sources magnetically the BI for $H = dB_2$. Similarly, the dilaton $\phi$ is dual to an $8$-form, sourced by a seven-brane. It is then to be expected that seven-branes source the ${\cal J}_{ij}$ BI associated to the dilaton. Finally, the counting suggests that ${\cal J}$ corresponds to the source of a nine-brane. Since this BI is associated to a truncation ${\cal N} = 8 \to 4$ in the contexts of gauged supergravities, it is possible that such a truncation is produced by this source.

The sources in string theory are related by U-dualities. For example, the D-branes are related by T-dualities
\be
D_0 \leftrightarrow D_1 \leftrightarrow D_2 \leftrightarrow \dots
\ee
and these objects source the BI  (\ref{BIRR}). In IIB, the $D_7$ is S-dual to the $NS7$ which in turn is connected to other objects through T-dualities (we refer to \cite{deBoer:2010ud} for explanations on the notation)
\be
NS7 \leftrightarrow 6^1_3  \leftrightarrow \dots  \leftrightarrow 1^6_3 \leftrightarrow 0^7_3
\ee
This could be related to the BI (\ref{BI2}). In fact, following the logic $NS5 \to B_6 \leftrightarrow B_2 \to {\cal J}_{ijkl}$, we can think of a similar relation for the $NS7$: $NS7 \to A_8 \leftrightarrow A_0 \to {\cal J}_{ij}$.  The $1^6_3$ is connected through S-duality with a $1^6_4$, which is T-dual to  a $0^{(1,6)}_4$ in IIB, and has the nice property of being T-dual to a $0^{(1,6)}_4$ in IIA. This ``duality invariance'' might relate it to the BI (\ref{BI0}).

We stress that this is mere speculation, that must be explored further. We believe that this formulation of DFT with a relaxed strong constraint can give rise to the possibility of describing bound states of exotic branes that can't be described in supergravity.

\section{Conclusions and open problems}\label{conclusions}

We considered a flux formulation of DFT in which the fluxes are dynamical and field dependent. In this formulation, the gauge consistency constraints of the theory take the form of generalized quadratic constraints for the fluxes, that are known to admit solutions that violate the strong constraint \cite{Aldazabal:2011nj}. Building on previous constructions for a geometric formulation of DFT \cite{Siegel:1993th},\cite{WaldramR},\cite{Park},\cite{Geom}, we computed connections and curvatures on the double space, under the assumption that covariance is achieved up to the generalized quadratic constraints, rather than the strong constraint. Interestingly, this procedure gives rise to all the strong constraint-violating terms in the action, which are gauge invariant and appear systematically. This completes the original formulation of the theory \cite{Hull:2009mi}, incorporating the missing terms that allow to make contact with  half-maximal gauged supergravities \cite{Aldazabal:2011nj} containing duality orbits of non-geometric fluxes \cite{Dibitetto:2012rk}.

The consistency constraints are shown to be related to generalized BI that break down on the world-volume of (exotic) branes \cite{deBoer:2010ud}. We have speculated on the sources for the duality orbits of the BI, but this analysis deserves further investigation. For example, in \cite{Bergshoeff:2012jb} the universal T-duality representations for branes in different dimensions were classified, and it would be interesting to explore if these objects can be related to the BI discussed here. More generally, the quadratic constraints arising in U-duality invariant constructions \cite{SS Uduality}, \cite{Aldazabal:2013mya} should be sourced by U-duality orbits of branes. It would also be interesting to incorporate source terms in the action in a T-duality invariant way, such that the source terms appear naturally in the consistency constraints of the theory (in the form of tadpole cancelation conditions). This seems to require an extension of the generalized diffeomorphisms.

There is by now plenty of evidence that the strong constraint or section condition can be relaxed in duality covariant frameworks \cite{Aldazabal:2011nj},\cite{Dibitetto:2012rk},\cite{SS Uduality},\cite{Aldazabal:2013mya},\cite{BermanGEOM}. Transcending supergravity, this opens the door to seek for new truly double solutions to the equations of motion, or their associated supersymmetric killing-spinor equations. The T-duality invariance of the theory allows to build new T-fold-like solutions, like those of \cite{Hassler:2013wsa}, but more generally a relaxed strong constraint would allow to find solutions that lack a local interpretation from a supergravity point of view, in any global frame. By now, the only known solutions to the minimal constraints are of the SS type (this includes the strong constraint case in the decompactification limit) but we believe that other kind of compactifications will lead to new possibilities.

This truly double construction is interesting on its own and useful to describe non-geometry. However, it is still not clear whether a relaxation of the strong constraint in DFT describes a trustable limit of string theory. We plan to come back on these points in the future.

\bigskip

%%%%%%%%%%%%%%%%%%%%%%%%%%%%%%%%%%%%%%%%%%%%%%%%%%%%%%%%%%%%%%%%%%%%%%%%%%%%%%%%%%%%%%%%%%%%%%%
{\bf \large Acknowledgments} {We thank G. Aldazabal, M. Blau, J-P. Derendinger, M. Gra\~na, I. Jeon, K. Lee, H. Montani, J-H. Park, A. Rosabal  and M. Shigemori for enlightening discussions and correspondence. DG and DM thank CQuEST in Seoul and together with CN they thank IPhT-CEA in Paris for hospitality during early stages of this work. The work of DG was supported by the Swiss National Science Foundation and that of DM, CN and VP by CONICET and University of Buenos Aires.}

\appendix

%%%%%%%%%%%%%%%%%%%%%%%%%%%%%%%%%%%%%%%%%%%%%%%%%%%%%%%%%%%%%%%%%%%%%%%%%%%%%%%%%%%%%%%%%%%%%%%
\section{$O(D,D)$ spinors}\label{oddspin}

Let $\{\G^A\} = \{\G^a, \G_a\}$ be a set of gamma matrices giving a representation of the Clifford algebra
\be\label{oddscliff}
	\left\{\G^A, \G^B \right\} = \eta^{AB},
\ee
defined here with a non-standard normalization, where $\eta^{AB}$ is the off-diagonal $O(D,D)$ metric. With this particular signature, the gamma matrices can always be chosen to be real, with the property
\be\label{oddstrans}
	(\G^a)^T = \G_a.
\ee
The charge conjugation matrix is then
\be\begin{split}\label{oddscc}
	C &= \left(\G_0-\G^0\right)\dots\left(\G_{D-1}-\G^{D-1}\right), \\
	(C \G^{(n)})^T &=  (-1)^\frac{(D-n)(D - n + 1)}{2} C \G^{(n)},
\end{split}\ee
where $\G^{(n)}$ is an antisymmetrized product of $n$ gamma matrices. For this signature, one can always impose a Majorana condition on spinors, which is a reality condition on spinors for real gamma matrices. Moreover, since the dimensionality is even, a chirality condition can be imposed with the product of all gamma matrices
\be
	\G^\ast = \left(1-2\G^0\G_0 \right)\dots \left(1-2\G^{D-1} \G_{D-1} \right),
\ee
where an ordering sign has been included. These matrices give a representation of a fermionic oscillators algebra, $\{\G_a,\G^b\} = \d_a^b$ and $\{\G_a,\G_b\} = \{\G^a,\G^b\} = 0$. A Clifford vacuum $\vert 0 \rangle$, normalized to $\langle 0\vert 0\rangle = 1$ and annihilated by $\G_a$, can then be defined. A (anti)chiral spinor is then obtained by acting on this vacuum with an (odd) even number of raising operators
\be
	\vert\omega\rangle
	=  \sum_k \frac{\omega_{a_1\dots a_k}}{k!}\G^{a_1}\dots \G^{a_k}\vert 0\rangle,
\ee
thus giving a map between a polyform $\omega = \sum_k \omega_{(k)}$ and a spinor $\vert\omega\rangle $. Using the charge conjugation matrix, an $O(D,D)$ invariant bilinear can be constructed. Up to a sign, it corresponds to the Mukai paring of two polyforms (the $D$-form in the product $\chi\wedge\sigma\,\omega$)\footnote{$\sigma$ is an operator reversing the order of the differentials $dx^i$ in a form.} and reads in components
\be
	\langle\chi\vert C\vert\omega\rangle
	= \sum_k \frac{(-1)^{k}}{k!(D-k)!}
		\epsilon^{a_1\dots a_D} \chi_{a_1\dots a_k}\omega_{a_D\dots a_{k+1}},
\ee
where $\vert \chi\rangle^T = \langle \chi\vert$. In order to define $H=O(1,D-1)\times O(1,D-1)$ invariant products, one needs to define the $Spin(D,D)$ representative of the metric $S_{AB}$, viewed as an $O(D,D)$ element, by
\be\begin{split}
	\Psi_\pm &=
	\left(\G_0 \mp \G^0\right)\left(\G_1 \pm \G^1\right)
	\dots\left(\G_{D-1} +\pm \G^{D-1}\right), \\
	\Psi_\pm^T &= \Psi_\pm^{-1} = -(-)^{\frac{D(D\mp 1)}{2}}\Psi_\pm.
\end{split}\ee
These matrices satisfy the following (anti-)commutation relation with a gamma matrix
\begin{equation}\label{ddemb_hgcomm}
	\Psi_\pm \G^A = \mp(-)^D S^{AB}\G_B \Psi_\pm,
\end{equation}
and are indeed spin representatives of $\mp(-)^D S^{AB}$ respectively, hence commuting with $H$-restricted spin transformations. Acting on the spinor $\vert\omega\rangle$ with $\Psi_+$ yields in components
\be
	\Psi_+ \vert\omega\rangle =
	\sum_k \frac{\epsilon_{a_1\dots  a_{(D-k)}}{}^{a_{(D-k+1)}\dots  a_D}}{k!(D-k)!}\,
	\omega_{a_D\dots a_{(D-k+1)}}\,\G^{a_1}\dots \G^{a_{(D-k)}}\vert 0 \rangle
	= \vert\star\sigma\,\omega\rangle,
\ee
where, in our conventions, the Hodge star is defined as
\be
	\star \omega_{(k)} = \frac{\sqrt{\vert g\vert}}{k!(D-k)!}
	\epsilon_{i_1\dots i_D}\, dx^{i_1}\wedge \dots \wedge  dx^{i_{D-k}} \,
	\omega^{i_{D-k+1} \dots  i_D},
\ee
and is pseudo-involutive $\star^2\omega_{(p)} = (-1)^{t + (D-p)p}$, for $t$ time-like directions. The $H$-invariant bilinear formed with $\Psi_+$ then reads in components
\be
	\langle\chi\vert C\Psi_+ \vert \omega\rangle
	= \sum_k \frac{s^{a_1 b_1}\dots s^{a_k b_k}}{k!} \chi_{a_1\dots a_k}\omega_{b_1\dots b_k}.
\ee
To make contact with the language of Generalized Complex Geometry, is possible to introduce curved gamma matrices $\G^M$. Note that, since the constrained bein $\cE_A{}^M$ is an element of $O(D,D)$, it is possible to choose at the same time $\G^M$ and $\G^A$ as constant matrices related by
\be\label{curvedgamma}
	S_\cE\, \G^M\, S^{-1}_\cE =  \G^A\,\cE_A{}^M,
\ee
where $S_\cE$ is the $Spin(D,D)$ representative of the bein. The derivative of this object is given by
\be\label{rr_srder}
    \cD_A S_\cE = -\frac{1}{2} \Omega_{ABC} S_\cE \G^{BC},
\ee
as found by asking compatibility with (\ref{curvedgamma}).

\section{General Relativity and anholonomy}\label{GR}

Let $d_a = e_a{}^i \p_i$ be a frame and $e^a = e^a{}_i dx^i$ its dual, where $e^a{}_i e_a{}^j = \delta_i^j$. Their respective structure equations read
\be
	\left[d_a,d_b \right]=\tau_{ab}{}^c\, d_c,
	\qquad de^a = -\frac{1}{2} \tau_{bc}{}^a\,e^b\wedge e^c\, ,
\ee
where the anholonomy coefficients read $\tau_{ab}{}^c = 2\Gamma_{[ab]}{}^c$, with $\Gamma_{ab}{}^c = \left(d_a e_b{}^i\right)e^c{}_j$ as defined in (\ref{Gammas}). These coefficients measure the failure of the frame to be locally a coordinate basis, i.e. $e^a = dy^a$. Taking the exterior derivative of the second structure equation yields
\be\label{mcbi}
	d_{[a}\, \tau_{bc]}{}^d + \tau_{[ab}{}^e\tau_{c]e}{}^d =0.
\ee
Contracting the upper index with one lower index also yields
\be\label{mcbi2}
	d_c \tau_{ab}{}^c + 2d_{[a} \tau_{b]c}{}^c - \tau_{ab}{}^c \tau_{cd}{}^d  = 0.
\ee
A covariant derivative $\nabla = d + \omega$ is introduced, the connection one-form acting on Lorentz indices as $\omega f^a = \omega^a_b f^b$ and $\omega f_a = -\omega^b_a f_b$. In the  $e^a$ basis it reads $\nabla_a f^b = e_a f^b + \omega^b{}_{ac}f^c$. The torsion two-form is defined as
\be\label{torsion}
	T^a = \nabla e^a = de^a + \omega^a{}_b\wedge e^b
	= \left(\omega^a{}_{bc} -\frac{1}{2}\tau_{bc}{}^c\right)\, e^b\wedge e^c.
\ee
The antisymmetric part of the spin connection is then fully determined in term of the torsion and anholonomy coefficients $\omega^c{}_{[ab]} = \frac{1}{2}\left(\tau_{ab}{}^c + T_{ab}{}^c\right)$. Asking for consistency with partial integration
\be
	\int d^Dx\ e\ U^{a_1 a_2\dots a_n}\nabla_{a_1}T_{a_2\dots a_n}
	= -\int d^Dx\ e\ T_{a_2\dots a_n}\nabla_{a_1}U^{a_1 a_2\dots a_n},
\ee
where $e = \det e^a{}_i$, further constrains one trace of the spin connection $\omega^b{}_{ba} = - \tau_{ab}{}^b $. The curvature two-form is defined as
\be\label{curvature}
	\nabla^2 f^a = R^a{}_b f^b
	= \left(d\omega^a{}_b + \omega^a{}_c\wedge\omega^c{}_b\right) f^b
	= \frac{1}{2}R^a{}_{bcd} \, e^c\wedge e^d\, f^b\, .
\ee
Taking the covariant derivative of (\ref{torsion}), one obtains
\be
	\nabla T^a = \frac{1}{2} R^a{}_{[bcd]}\, e^b\wedge e^c\wedge e^d,
\ee
such that vanishing torsion implies the cyclic Bianchi identity for the Riemann tensor. In the zero-torsion case, the antisymmetric part $R^a{}_{[bcd]}$ only depends on the antisymmetric part of the spin connection $\omega^c{}_{[ab]} = \frac{1}{2} \tau_{ab}{}^c$ and vanishes due to identity (\ref{mcbi})
\be
	R^d{}_{[abc]} = d_{[a}\, \tau_{bc]}{}^d + \tau_{[ab}{}^e\tau_{c]e}{}^d = 0.
\ee
The Ricci tensor in the $e^a$ basis reads
\be
	R_{ab} = R^c{}_{acb}.
\ee
Using the relation $\omega^b{}_{ba} = - \tau_{ab}{}^b $, its antisymmetric part vanishes due to identity (\ref{mcbi2})
\be
 R_{[ab]} =
 \frac{1}{2}\left( d_c \tau_{ab}{}^c - d_b \tau_{ac}{}^c
 + d_a \tau_{bc}{}^c - \tau_{cd}{}^d \tau_{ab}{}^c\right) = 0.
\ee
Introducing the Lorentz metric $s_{ab}$, a metric compatibility condition can be imposed on the connection by asking the metric to be covariantly constant $\nabla s_{ab} = -2\omega_{(ab)} = 0$. For vanishing torsion, this condition is solved by \be
	\omega^c{}_{ab} = \frac{1}{2}\tau_{ab}{}^c + \tau^c{}_{(ab)},
\ee
where indices on the l.h.s. are raised and lowered with the Lorentz metric $s_{ab}$. For this choice of connection the Ricci tensor reads
\bea
	R_{ab} &=& d_{(a}{{\tau}_{b) c}{}^{c}}
	+ d_{c}{{\tau}^{c}{}_{(a b)}}
	- \frac{1}{4}\, {\tau}_{a c}\,^{d} {\tau}_{b d}\,^{c}
	+ \frac{1}{2}\, {\tau}_{a c}\,^{d} {\tau}^{c}\,_{(b d)}
	+ \frac{1}{2}\, {\tau}_{b c}\,^{d} {\tau}^{c}\,_{(a d)} \\
	&&- {\tau}_{dc}\,^{c} {\tau}^{d}\,_{(a b)}
	- \frac{1}{2}\, {\tau}^{c}\,_{a d} {\tau}^{d}\,_{(b c)}
	- \frac{1}{4}\, {\tau}^{c}\,_{b d} {\tau}^{d}\,_{c a}
	- \frac{1}{4}\, {\tau}^{c}\,_{d a} {\tau}^{d}\,_{c b},
\eea
and the Ricci scalar can be computed
\be
	R = s^{ab}\, R_{ab}
	= 2d_a\, \tau^{ab}{}_b -\tau^{ab}{}_b\tau_{ac}{}^c
	-\frac{1}{2}\tau^a{}_b{}^c\tau_{ac}{}^b -\frac{1}{4}\tau^{ab}{}_c\tau_{ab}{}^c.
\ee

%%%%%%%%%%%%%%%%%%%%%%%%%%%%%%%%%%%%%%%%%%%%%%%%%%%%%%%%%%%%%%%%%%%%%%%%%%%%%%%%%%%%%%%%%%%%%%%

\end{document}